\input{epsf}
\documentstyle[eqsecnum,aps]{revtex}
\tightenlines
\begin{document}

\title{Energy-Momentum Tensor of Particles Created in an Expanding Universe}

\vspace{1.5 true cm}

\author{Salman Habib
\thanks{electronic address: habib@lanl.gov},
Carmen Molina-Par\'{\i}s
\thanks{electronic address: carmen@t6-serv.lanl.gov},
and Emil Mottola
\thanks{electronic address: emil@lanl.gov}}

\address{Theoretical Division T-8, Los Alamos
National Laboratory, Los Alamos, New Mexico, 87545}

\date{\today}

\preprint{LA-UR-99-2812}

\maketitle

\begin{abstract}
\baselineskip 1 pc

We present a general formulation of the time-dependent initial value
problem for a quantum scalar field of arbitrary mass and curvature
coupling in a Friedmann-Robertson-Walker (FRW) cosmological model. We
introduce an adiabatic number basis which has the virtue that the
divergent parts of the quantum expectation value of the
energy-momentum tensor $\langle T_{a b}\rangle$ are isolated in the
vacuum piece of $\langle T_{a b}\rangle$, and may be removed using
adiabatic subtraction. The resulting renormalized $\langle
T_{ab}\rangle$ is conserved, independent of the cutoff, and has a
physically transparent, quasiclassical form in terms of the average
number of created adiabatic `particles'.  By analyzing the evolution
of the adiabatic particle number in de Sitter spacetime we exhibit the
time structure of the particle creation process, which can be
understood in terms of the time at which different momentum scales
enter the horizon. A numerical scheme to compute $\langle T_{ab}
\rangle$ as a function of time with arbitrary adiabatic initial states
(not necessarily de Sitter invariant) is described. For minimally
coupled, massless fields, at late times the renormalized $\langle
T_{ab}\rangle$ goes asymptotically to the de Sitter invariant state
previously found by Allen and Folacci, and not to the zero mass limit
of the Bunch-Davies vacuum.  If the mass $m$ and the curvature
coupling $\xi$ differ from zero, but satisfy $m^2+\xi R=0$, the energy
density and pressure of the scalar field grow linearly in cosmic time
demonstrating that, at least in this case, backreaction effects 
become significant and cannot be neglected in de Sitter spacetime.
\thispagestyle{empty}
\end{abstract}

\voffset=1.0 true cm

\newpage

\voffset=-0.5 true cm

\pagestyle{plain}

\pagenumbering{arabic}

\newcommand{\sq}{\lower.15ex\hbox{\large$\Box$}}

\section{Introduction}

\label{sec:intro}

Quantum particle creation effects are important in black hole
physics~\cite{blackhole1,blackhole2}, the generation of density
perturbations and primordial gravitational waves~\cite{gravitywaves},
and reheating in inflationary cosmological models~\cite{reheating},
the damping of anisotropy in the early Universe~\cite{damping}, and
perhaps for even the very existence of a cosmological
constant~\cite{cosmoconstant,sahni-habib}. In addition, the issue of quantum
backreaction due to these effects could also be important for
understanding black holes~\cite{br-blackhole} and the dynamics of the
early Universe~\cite{br}.

In a curved spacetime background, it is not possible in general to
define a unique vacuum state~\cite{semiclassical}. However, if the
background expansion is sufficiently slow the notion of an adiabatic
vacuum state defined via WKB expansions comes closest to the
Minkowski vacuum~\cite{books}. For FRW cosmological models, adiabatic
regularization~\cite{adiabatic} has been shown to be particularly
convenient, and equivalent to covariant
point-splitting~\cite{equivalence,anderson-parker,christensen}.

In this paper we present a consistent, physically simple formulation
of the general time-dependent initial value problem for a quantum
scalar field of arbitrary mass $m$, and curvature coupling $\xi$ in a
background homogeneous, isotropic FRW cosmology, which is suitable for
implementation on a computer. We present numerical results for
particle creation in de Sitter space, and outline how the formalism
can be applied to cosmological backreaction.

We make use of an adiabatic particle number basis obtained via a
time-dependent Bogoliubov transformation, in which the energy-momentum
tensor of a quantum field in a FRW universe breaks up naturally into a
`particle' and a `vacuum' contribution, the latter isolating the
ultraviolet divergences. These divergences are then removed using
adiabatic subtractions. The remaining finite energy density and
pressure of the quantum field admit an intuitive quasiclassical
interpretation in terms of the average number of particles present at
a given time. These terms can be interpreted as the energy density and
pressure of the produced `particles' if the Universe entered an
adiabatic era. This definition is motivated by and is a direct
outgrowth of parallel definitions of particle number in time dependent
problems in field theories in Minkowski space, such as scalar QED and
$\lambda \Phi^4$ theory~\cite{diss,phase,kluger}.

It is well known that the concept of particle is ambiguous in time
dependent backgrounds. Covariantly conserved quantities such as the
renormalized expectation value of $\langle T_{ab} \rangle$ can be
defined without ambiguity (save the usual freedom associated with the
additive finite renormalization of conserved local tensors), although
the division of $\langle T_{ab} \rangle$ into `vacuum' and `particle'
components is not unique. However, if one requires that the
intrinsically quantum ultraviolet power divergences of $\langle T_{ab}
\rangle$ are isolated in the `vacuum' component of $\langle T_{ab}
\rangle$, then the ambiguity in the remainder is considerably
reduced. In fact, this requirement is equivalent to matching the
definition of `vacuum' to an adiabatic basis up to at least second
adiabatic order. One does not wish to match the vacuum to higher than
fourth adiabatic order since this would introduce higher than fourth
order time derivatives in the backreaction equations. Hence the only
remaining ambiguity in the definition of the `particles' resides in
the treatment of the fourth adiabatic order terms, and is actually
relatively mild. We introduce a precise definition of adiabatic basis
in this paper which resolves the remaining ambiguity and use it to
renormalize the energy-momentum tensor and study the particle creation
process in de Sitter background. The resulting renormalized $\langle
T_{ab}\rangle$ has a physically transparent, quasiclassical form in
terms of the average number of created `particles', and is the
appropriate source term for dynamical backreaction of the quantum
field on the metric in semiclassical cosmology. Since our main
interest here is the initial value problem in semiclassical gravity,
we consider general FRW-invariant initial states. To prevent the
occurrence of initial time singularities, we restrict attention to
initial conditions which are also matched to the appropriate
instantaneous fourth order adiabatic vacuum at the initial
time~\cite{lindig}.

The renormalized $\langle T_{a b}\rangle$ we obtain is cutoff
independent, and covariantly conserved. As a consequence of our
analysis we demonstrate that these features do not depend on the
choice of time parametrization ({\it e.g.,} comoving versus conformal
time), thus clarifying some previous discussions in the
literature~\cite{boyan}. We have implemented a numerical scheme to
compute the time evolution of the renormalized $\langle T_{ab}\rangle$
which preserves covariant conservation and cutoff independence. Though
our scheme is applicable to general FRW spacetimes, in this paper we
have restricted ourselves to de Sitter space with closed spatial
sections. We have checked covariant conservation and cutoff
independence of the renormalized $\langle T_{ab}\rangle$ for a variety
of masses, couplings to curvature, spatial cutoffs, and non de
Sitter-invariant initial states.

We consider in some detail the time structure of the particle creation
process for massive scalar particles in de Sitter space starting from
a well-defined adiabatic initial state, which is not de Sitter
invariant. The particle number in a particular momentum mode increases
sharply when the physical wavelength of that mode crosses the de
Sitter horizon. We show how the adiabatic particle basis is useful in
interpreting the numerical results for the time evolution of the
renormalized energy and pressure. We apply the same method to the
massless case with $\xi=0$ and $\xi=1/6$. For the case $\xi=1/6$, as
expected from conformal invariance, there is no particle production,
and the renormalized value of $\langle T_{ab}
\rangle$ is determined by the trace anomaly~\cite{bunch-davies}.

The case of $M^2 = m^2 + \xi R = 0$ is important because the two
helicity states of gravitons obey a wave equation which corresponds to
this value of the mass parameter $M^2$, and there are several
indications of strong infrared effects for such fields in de Sitter
space. For example it is known that no de Sitter invariant state of
the kind usually assumed, implicitly or explicitly, in inflationary
models of the early universe exists for massless, minimally coupled
scalar fields~\cite{folacci}, and that the graviton propagator has
unusual infrared behavior in de Sitter
space~\cite{Antoniadis-Mottola}.

We will see that the exact minimally coupled, massless case $m^2 =
\xi = 0$ is special in that in this case, at late times the
renormalized $\langle T_{ab}\rangle$ goes asymptotically to the
constant de Sitter invariant value found by Allen and
Folacci~\cite{folacci}, and not to the zero mass limit of the
Bunch-Davies vacuum~\cite{bunch-davies}.  For any other nonzero values
of $m^2$ and $\xi$ such that $M^2 = 0$ the energy-momentum tensor
{grows without bound}, linearly with cosmic time $t$, for adiabatic
finite initial data. (We point out that this result is a quantum
effect, and not a classical instability.)  Hence at least in these
cases it is clear that the backreaction of $\langle T_{ab}\rangle$ on
the metric cannot be neglected at late times. We have also
investigated the case of a small mass, minimally coupled field,
results for which will be reported elsewhere~\cite{smallm}.

The outline of the paper is as follows. In the next section we review
scalar quantum field theory in general FRW spacetimes, which will
serve to fix our notation and conventions. In Section III we introduce
the adiabatic number basis by a certain time dependent Bogoliubov
transformation on the exact mode functions. In Section IV we derive
the energy density and pressure for a scalar field of arbitrary mass
and curvature coupling in homogeneous, isotropic cosmologies as sums
(or integrals) over the time dependent Fourier mode amplitudes and
express the energy and pressure in the adiabatic number basis, showing
explicitly how to isolate and remove the ultraviolet divergences in
the vacuum contributions to these mode sums. In Section V we present
numerical results for de Sitter space, for the massive, conformally
coupled and minimally coupled scalars, for a variety of finite non de
Sitter invariant initial states. In Section VI we present our
conclusions and discuss the extension of our method to the full
backreaction problem. A proof of the independence of the adiabatic
method under time reparametrizations is given in the Appendix.

\section{Scalar Field in FRW Spacetimes}

We consider spatially homogeneous and isotropic spacetimes which are
described by the line element
\begin{eqnarray}
{\mathrm{d}}s^2= -{\mathrm{d}}
t^2 + g_{ij}{\mathrm{d}}x^i {\mathrm{d}}x^j = - {\mathrm{d}}t^2 
+ a^2(t)\, {\mathrm{d}}{\Sigma}^2
\; ,
\label{FRWs}
\end{eqnarray}
in the cosmic time $t$, comoving with freely falling observers.  Here
${\mathrm{d}}\Sigma^2$ is the line element of spatial sections with constant
curvature, which can be labeled by $\kappa = -1,0,+1$, {\it viz.},
\begin{eqnarray}
{\mathrm{d}} \Sigma^2 = {{\mathrm{d}}r^2 \over 1 -\kappa r^2}
+ r^2 ({\mathrm{d}}\theta^2 + \sin^2\theta \,{\mathrm{d}}\varphi^2)\,.
\label{spatial}
\end{eqnarray}
The conformal time variable $\eta$ is defined by the differential
relation
\begin{eqnarray}
{\mathrm{d}}\eta = {{\mathrm{d}}t \over a(t)}
\; ,
\label{conftim}
\end{eqnarray}
and the function
\begin{eqnarray}
C(\eta) \equiv a^2\left(t(\eta)\right)\,,
\label{Cdef}
\end{eqnarray}
so that the line element (\ref{FRWs}) may be expressed in the
equivalent form,
\begin{eqnarray}
{\mathrm{d}}s^2= C(\eta)\ (- {\mathrm{d}}\eta^2 + {\mathrm{d}}\Sigma^2)\,.
\end{eqnarray}
It is the latter form that was considered by Bunch~\cite{Bunch}, a
reference to which some of our results in this and the next section
may be compared.

The time dependent Hubble function in comoving time is
\begin{eqnarray}
H(t) \equiv {{\dot a(t)} \over {a(t)}} = {1\over 2} {{\mathrm{d}}\over 
{\mathrm{d}}t} \log C
\; ,
\end{eqnarray}
where the dot denotes differentiation with respect to comoving time
$t$.  The non-vanishing components of the Riemann, Ricci, and Einstein
tensors, $G_{ab} = R_{ab} - {1 \over 2} R g_{ab}$, are given by
\begin{eqnarray}
R_{itjt} &=& R_{titj} = -(\dot H + H^2) g_{ij}\,,\\
R_{ijkl} &=& \left( H^2 + {\kappa \over a^2}  \right) 
(g_{ik}g_{jl} - g_{il}g_{jk})\,,\\ 
R_{tt} &=& - 3 (\dot H + H^2)\,,\\ 
R_{ij}&=&  \left(\dot H + 3 H^2 + {{2 \kappa} \over a^2}  \right) g_{ij}\,,\\
G_{tt} &=& 3 \left( H^2 + {\kappa \over a^2}  \right)\,,\\ 
G_{ij}&=& - \left( 2 \dot H + 3 H^2 + {{ \kappa} \over a^2}  \right) g_{ij}\,.
\end{eqnarray}
Hence the the scalar curvature is
\begin{eqnarray}
R &=& g^{ab}R_{ab} = - R_{tt} + g^{ij}R_{ij} 
= 6 \left[ \dot H(t) + 2
H^2(t) + {\kappa \over a^2}\right]\,. 
\end{eqnarray} 
We will require the non-zero components of the conserved geometric
tensor,
\begin{eqnarray}
^{(1)}H_{ab} &\equiv& {1\over \sqrt{-g}} {\delta\over\delta g^{ab}}
\int\, \sqrt{-g}\,{\mathrm{d}}^4 x R^2
\nonumber \\
&=& 2\nabla_a\nabla_b R - 
2 g_{ab} \sq R -{1\over 2} g_{ab} R^2 + 2 R R_{ab}\,,
\label{Hone}
\end{eqnarray}
as well, which are given by
\begin{eqnarray}
^{(1)}H_{tt} &=& -6 H \dot R + {1\over 2} R^2 -6 (H^2 +\dot H) R\,,
\nonumber
\\
^{(1)}H_{ij} &=& \left[- 2 \sq R - 2 H \dot R + {1\over 6} R^2 
-2 (H^2 +\dot H) R\right]g_{ij}\,,
\nonumber\\
^{(1)}H_a^a &=& -6\sq R = 6 (\ddot R + 3H\dot R)\,.
\label{Honecomp}
\end{eqnarray}
In a conformally flat FRW spacetime the Weyl tensor, defined as
follows
\begin{eqnarray}
C_{abcd} &\equiv& R_{abcd} - {1\over 2} \left( g_{ac}R_{bd} + g_{bd}R_{ac}
-g_{ad}R_{bc}-g_{bc}R_{ad}\right) 
+ {1\over 6} R \left(g_{ac}g_{bd}-
g_{ad}g_{bc}\right) 
\, ,
\end{eqnarray}
vanishes. From this it follows that the geometric tensor
\begin{eqnarray}
^{(3)}H_{ab} = R_a^cR_{cb} -{2\over 3}RR_{ab} -{1\over 2} R_{cd}R^{cd}g_{ab}
+ {1\over 4} R^2 g_{ab}
\nonumber
\; ,
\end{eqnarray}
which satisfies
\begin{eqnarray}
\nabla^b\ ^{(3)}H_{ab} = 2 R^{cd} \nabla_b C^b_{cda} =0
\; ,
\end{eqnarray}
is covariantly conserved in conformally flat FRW spacetimes as well.

We consider in this paper a free scalar field $\Phi$ with arbitrary
mass and curvature coupling, which is described by the quadratic
action,
\begin{eqnarray}
S = - {1\over 2}\int\,\sqrt{-g}\ 
{\mathrm{d}}^4 x 
\, \left[(\nabla_a \Phi)g^{ab}
(\nabla_b \Phi) 
+ m^2\Phi^2  +\xi R \Phi^2\right]
\; ,
\label{action}
\end{eqnarray}
in a general curved spacetime, where $\nabla_a$ denotes the covariant
derivative and $g = \det (g_{ab})$. The wave equation for $\Phi$
obtained by varying this action is
\begin{eqnarray}
\left[- \sq + m^2 + \xi R\right] \Phi(t, {\bf x}) = 
\left[- \sq + M^2\right] \Phi(t, {\bf x})= 0\,. \label{waveq}
\end{eqnarray}
In a FRW spacetime expressed in comoving time coordinates (\ref{FRWs})
the D'Alembert wave operator on scalars becomes
\begin{eqnarray}
-\sq \Phi = -g^{ab} \nabla_a \nabla_b \Phi = \ddot \Phi + 3 H \dot \Phi 
-\frac{1}{a^2} \Delta_3 \Phi\,,
\end{eqnarray}
where $\Delta_3$ is the Laplace-Beltrami operator in the three
dimensional spacelike manifold of constant curvature $\kappa$.  This
operator is diagonalized by the spatial harmonic functions $Y_{\bf
k}({\bf x})$ which satisfy
\begin{eqnarray}
-\Delta_3 Y_{\bf k}({\bf x}) = (k^2 - \kappa)Y_{\bf k}({\bf x})\,,
\label{eigen}
\end{eqnarray}
where $k\equiv \vert{\bf k}\vert$ takes on the discrete values of the
positive integers $1, 2, \dots$ in the case of $\kappa = 1$, but
becomes a continuous index over the non-negative real numbers in the
case of $\kappa = 0, -1$. In the case of flat spatial sections,
$\kappa = 0$, the harmonic functions $Y_{\bf k}({\bf x})$ become the
ordinary Fourier plane wave modes $e^{i {\bf k \cdot x}}$. In the
compact case the harmonic functions are the spherical harmonics of the
sphere $S^3$, which are labeled by three integers ${\bf k}
\rightarrow (k, \ell, m)$, the latter two of which refer to the
familiar spherical harmonics on $S^2$ with $\ell \le k-1$. Thus a
given eigenvalue of $\Delta_3$ labeled by $k$ is $k^2$-fold
degenerate.  We note that in the present conventions the lowest mode
is the $k=1$ mode which is the {\it constant} mode on $S^3$. The
scalar spherical harmonics $Y_{klm}$ may be normalized in the standard
manner to satisfy
\begin{eqnarray}
\int_{S^3} 
{\mathrm{d}}^3\Sigma \ Y^*_{k'\ell'm'}({\bf x})\, Y_{k\ell m}({\bf x}) =
\delta_{k'k} \delta_{\ell'\ell}\delta_{m'm}
\; ,
\end{eqnarray}
when integrated over the unit three sphere, and
\begin{eqnarray}
\sum_{\ell = 0}^{k-1}\sum_{m=-\ell}^{\ell} \vert Y_{k\ell m}({\bf x})\vert^2
= {k^2 \over 2\pi^2}\,,
\label{lmsum}
\end{eqnarray} 
which is independent of $\bf{x}$.

Because of homogeneity and isotropy the wave equation (\ref{waveq})
separates in either comoving or conformal time coordinates, and the
general solution of (\ref{waveq}) may be written in the form
\begin{eqnarray}
\Phi(t,{\bf x})= a^{-{3\over 2}}(t) \sum_{\bf k}
\; [ a_{\bf k}
f_k(t) Y_{\bf k}({\bf x}) 
+ a^{\dagger}_{\bf k} f^{*}_k(t) Y^*_{\bf k}({\bf x})]\,,
\label{modedec}
\end{eqnarray}
where the time dependent mode functions $f_k(t)$ satisfy the ordinary
differential equation,
\begin{eqnarray}
\ddot f_k(t) + \Omega^2_k(t) f_k(t)=0\,,
\label{modefns}
\end{eqnarray}
which is the equation of a harmonic oscillator with the time varying
frequency,
\begin{eqnarray}
\Omega_k^2 (t) = \omega_k^2 (t) + \delta \omega_k^{2} (t) \equiv
\left({k^2 \over a^2} + m^2\right) +
 \left(\xi - {1\over 6}\right) R(t) - {1 \over 2}
\left[ \dot H(t) + {H^2(t) \over 2} \right]\,.
\label{bigom}
\end{eqnarray}
We denote by $\omega_k^2$ the first two terms in parentheses, which
involve no time derivatives of the metric, while $\delta
\omega_k^{2}$
denotes the remaining terms which are second order in time derivatives
of the scale factor. In (\ref{modedec}) we use the discrete notation
or the sum $\sum_{\bf k}$, which is replaced by an integral over
${\mathrm{d}}^3{\bf k}$ in the cases $\kappa = 0, -1$.
  
Quantization of the scalar field is effected by requiring the equal
time canonical commutation relation
\begin{eqnarray}
[\Phi (t,{\bf x'}), \Pi (t,{\bf x})] = i\hbar \delta^3 ({\bf x'}, {\bf x})
\; ,
\end{eqnarray}
where $\Pi = \sqrt {-g} \dot\Phi = a^3 \dot \Phi$ is the momentum
conjugate to $\Phi$. Because of the completeness and orthonormality of
the spatial harmonic functions $Y_{\bf k} ({\bf x})$ this canonical
commutation relation is satisfied provided the creation and
destruction operators obey
\begin{eqnarray}
[a_{\bf k}, a^{\dagger}_{\bf k'}] = \delta_{{\bf k}{\bf k'}} = \delta_{k'k}
\delta_{\ell'\ell}\delta_{m'm}
\; ,
\end{eqnarray}
and the complex mode functions obey the Wronskian condition,
\begin{eqnarray}
\dot f_k f_k^* - f_k \dot f_k^* = -i\hbar\,,
\label{wron}
\end{eqnarray}
which they may be chosen to satisfy at the initial time $t_0$. From
the mode equation of motion (\ref{modefns}) this Wronskian condition
will be fulfilled then for all subsequent times.

We will consider states of the scalar field, which like the geometry
(\ref{FRWs}), are also spatially homogeneous and isotropic. This
implies that the expectation value of the particle number operator
\begin{eqnarray}
\langle a^{\dagger}_{\bf k}a_{\bf k}\rangle = 
\langle a_{\bf k}a^{\dagger}_{\bf k}\rangle - 1 = N_k
\; ,
\label{part}
\end{eqnarray}
can be a function only of the magnitude $|{\bf k}| = k$.  In the
context of Minkowski spacetime quantum field theory it has been shown
that it is always possible to fix the bilinears~\cite{diss}
\begin{eqnarray}
\langle a^{\dagger}_{\bf k}a^{\dagger}_{\bf k}\rangle = 
\langle a_{\bf k}a_{\bf k}\rangle = 0
\; ,
\end{eqnarray}
by making use of the freedom in the initial phases of the mode
functions $f_k$, with no loss of generality. Precisely the same
conclusion applies here. Because of the ubiquitous appearance of the
Bose-Einstein factor in succeeding sections we will often make use of
the the definition
\begin{eqnarray}
\sigma_k \equiv 2 N_k + 1 
\; ,
\end{eqnarray} 
in what follows. The `vacuum' state in this basis corresponds to the
choice $\sigma_k = 1$.

\section{Adiabatic Number Basis}

The observation underlying the introduction of the adiabatic basis is
that the mode equation (\ref{modefns}) generally possesses time
dependent solutions which have no clear {\em a priori} physical
meaning in terms of particles or antiparticles. The familiar notion
that positive energy solutions of the wave equation correspond to
particles while negative energy solutions correspond to antiparticles
is quite meaningless in time dependent backgrounds, where the energy
of individual particle/antiparticle modes is not conserved, and no
such neat invariant separation into positive and negative energy
solutions of the wave equation is possible. This is just a reflection
of the fact that particle number does not correspond to a sharp
operator which commutes with the Hamiltonian, {\em i.e.,}
particle/antiparticle pairs are created or destroyed, and physical
particle number is not conserved in time dependent
backgrounds~\cite{parker69}.

It is equally clear physically that in the limit of slowly varying
time dependent backgrounds there should be an appropriate slowly
varying particle number operator which becomes exactly conserved in
the limit of $\dot a(t)$ vanishing. Clearly this slowly varying
particle number is {\it not} the $N_k$ defined with respect to the
time independent Heisenberg basis by (\ref{part}) above. This $N_k$ is
part of the initial data, a strict constant of motion, no matter how
rapidly varying the spacetime geometry is. The physical particle
number at time $t$ must be defined instead with respect to a time
dependent basis which permits a semiclassical correspondence limit to
ordinary positive energy plane wave solutions in the limit of slowly
varying $\Omega_{k}(t)$. This adiabatic basis is specified by
introducing the adiabatic mode functions~\cite{parker69,zeldovich}
\begin{eqnarray}
\tilde f_{k}(t) &=& \sqrt{\hbar \over 2 W_k(t)}\,
{\rm exp}\left({-i\Theta_k(t)}\right)\,, 
\; \; \; {\rm with} \; \; \; 
\Theta_k(t) \equiv \int^t 
{\mathrm{d}} t'\, W_k(t')\ ,
\label{adbmod}
\end{eqnarray}
which are guaranteed to satisfy the Wronskian condition (\ref{wron})
for any real, positive function $W_k (t)$. The phase $\Theta_k(t)$ in
(\ref{adbmod}) can be measured from any convenient point such as the
initial time $t_0$.

If one requires $\tilde f_k(t)$ to be a solution of the {\it exact}
mode equation (\ref{modefns}), a second order non-linear equation for
the frequency $W^2_k = \Omega^2_k + {3\,\dot W^2_k \over 4\,W^2_k} -
\frac{\ddot W_k}{2\, W_k}$ is obtained. However, the utility of the
WKB form of this equation is that instead of solving for $W_k$
exactly, it can be used to generate an asymptotic series in order of
time derivatives of the background metric. Terminating this asymptotic
series at a given order defines an adiabatic mode function $\tilde
f_{k}(t)$ (to that order), which will no longer be exact solution of
the original mode equation (\ref{modefns}), but which can serve as a
template adiabatic basis against which the exact mode functions
$f_k(t)$ can be compared. This is accomplished by expressing the exact
mode functions in the adiabatic basis by means of a time dependent
Bogoliubov transformation of the general
form~\cite{parker69,zeldovich},
\begin{eqnarray}
f_k(t)= \alpha_k(t) \tilde f_k(t) + \beta_k(t) \tilde f_k^*(t)\,.
\label{bogf}
\end{eqnarray} 
The precise form of the time dependent Bogoliubov coefficients
$\alpha_k$ and $\beta_k$ will be determined once we specify completely
the adiabatic basis functions $\tilde f_k$, and these will be used to
define a time dependent adiabatic particle number.

The first few orders of the asymptotic adiabatic expansion of the
frequency $W_k$ in powers of the time variation of the metric are
given by
\begin{eqnarray} W^{(0)\,2}_k &=& 
\omega^2_k \equiv
{k^2 \over a^2} + m^2\,,
\nonumber\\ 
W^{(2)\,2}_k &=& 
\Omega^2_k + {3\,\dot
W^{(0)2}_k \over 4\,W^{(0)\,2}_k} 
- {\ddot W^{(0)}_k \over 2\,W^{(0)}_k}=
\Omega^2_k + {3\,\dot \omega^2_k \over 4\,\omega^2_k} 
- {\ddot \omega_k \over
2\,\omega_k}\,,
\nonumber\\ 
W^{(4)\,2}_k &=& \Omega^2_k + {3\,\dot W^{(2)2}_k
\over 4\,W^{(2)\,2}_k} - {\ddot W^{(2)}_k \over 2\,W^{(2)}_k}\, ,
\qquad {\rm etc}\, .
\label{expand}
\end{eqnarray}
Explicitly to second adiabatic order in the asymptotic expansion,
\begin{eqnarray}
W^{(2)}_k = \omega_k + {\left(\xi - {1\over 6}\right) \over 2\,\omega_k}R
- {m^2\over 4\,\omega_k^3}(\dot H + 3H^2) + {5\,m^4 \over 8\, \omega_k^5} H^2
+ \dots \,,
\label{adbtwo}
\end{eqnarray}
where we have dropped third and higher order terms in the ellipsis.
{From} this we evaluate
\begin{eqnarray}
-{\dot W^{(2)}_k\over W^{(2)}_k } &=&
 H \left(1 - {m^2\over \omega_k^2}\right) 
- {\left(\xi - {1\over 6}\right)\over 2\,\omega_k^2} 
(\dot R + 2 R H) 
\label{adbthree}\\
&+& {\left(\xi - {1\over 6}\right) m^2\over \omega_k^4} R H
+ {m^2 \over 4 \omega_k^4}(\ddot H + 52 \dot H H + 18 H^3)
- {11\, m^4 \over 4\,\omega_k^6}(4\dot H + 3H^2)H 
+ {15\,m^6 \over 4\,\omega_k^8} H^3 +\dots
\,,\nonumber
\end{eqnarray}
correct to adiabatic order three.  Finally we have
\begin{eqnarray}
\label{adbfour}
W^{(4)}_k &=&  
W^{(2)}_k - {\left(\xi - {1\over 6}\right)^2 \over 8\,\omega_k^3}
\, R^2 - {\left(\xi - {1\over 6}\right) \over 8\,\omega_k^3}
\,(\ddot R +5\dot R H + 2R\dot H +6 RH^2)
\\
&+& {\left(\xi - {1\over 6}\right) m^2\over 8\,\omega_k^5}\, 
(5 \dot R H + 3 R \dot H + 19 R H^2)
+ {m^2\over 16\,\omega_k^5}\, 
({\stackrel{\dots}{H}} + 15 \ddot H H + 10 \dot H^2 
+ 86 \dot H H^2 + 60 H^4)
\nonumber\\
&-& {25 \left(\xi - {1\over 6}\right) m^2 \over 16\,\omega_k^7}\, 
R H^2 - {m^4\over 32\,\omega_k^7}\, 
(28 \ddot H H + 19 \dot H^2 + 394 \dot H H^2 + 507 H^4)
\nonumber\\
&+& {221\,m^6\over 32\,\omega_k^9}\, (\dot H + 3 H^2)H^2
- {1105\,m^8\over 128\,\omega_k^{11}}\, H^4 + \dots \, ,
\nonumber
\end{eqnarray}
where terms of adiabatic order higher than four have been
discarded. Notice that the terms fall off faster with large $k$ as we
develop the expansion to higher adiabatic orders, and therefore the
adiabatic expansion is an asymptotic expansion in powers of $1/k$
which matches the behavior of the exact mode functions in the far
ultraviolet region. This is what makes the expansion useful for
defining a particle number basis, since it can be used to isolate the
divergences of the vacuum energy-momentum tensor. Since ${\hat
T}_{ab}$ is a dimension four operator in four spacetime dimensions,
its divergences are obtained by expanding to adiabatic order four, and
no higher orders are required. In fact all terms in $W_k^{(4)}$ beyond
the first three of (\ref{adbfour}) fall off faster than
$\omega_k^{-3}$ as $k \rightarrow \infty$, and therefore will lead to
convergent mode sums (integrals) and finite contributions to $\langle
T_{ab}\rangle$.

Our method will make use of these well-known results and take them one
step further. Instead of demanding that $\tilde f_{k}(t)$ satisfy the
exact mode equation (\ref{modefns}) we will {\it define} $W_k(t)$ to
match the terms in $W^{(4)}_k$ that fall off most slowly in $k$, in
order to isolate the divergences of the energy-momentum tensor in the
vacuum sector in the most convenient way. The exact mode functions
$f_{k}(t)$ will be written then as a linear combination of $\tilde
f_{k}(t)$ defined by (\ref{adbmod}) and its complex conjugate. The
$\tilde f_{k}(t)$ defined in this way specify an adiabatic number
basis in which particle creation in the homogeneous, isotropic state
of the quantum field $\Phi$ can be discussed in a physically
meaningful way, free of ultraviolet divergences.

Since the Bogoliubov transformation (\ref{bogf}) may be viewed as a
pair of canonical transformations in the phase space spanned by the
field variables and their conjugate momenta, a complete specification
of the transformation between bases requires a condition on the first
time derivatives of the mode functions as well. The general form of
this condition which preserves the Wronskian relation (\ref{wron}) is
\begin{eqnarray}
\dot f_k(t)= \left[-i W_k(t) + {V_k(t) \over 2} \right] 
\alpha_k(t) \tilde f_k(t) 
+  \left[ i W_k(t) + {V_k(t) \over 2} \right] \beta_k(t) \tilde f_k^*(t)\,,
\label{dbog}
\end{eqnarray} 
which introduces a second real function of time $V_k$. Usually this
second independent function $V_k$, which contains only {\it odd}
adiabatic orders, has been set equal to $-\dot W_k /W_k$ in earlier
discussions of the adiabatic expansion. Hence the general adiabatic
basis as a canonical transformation in phase space, where $W_k$ and
$V_k$ are treated independently and on an equal footing, has not been
fully realized. We will make use of the freedom to choose both $W_k$
and $V_k$ independently in our development.  The precise choice of
$W_k$ and $V_k$ will be postponed until the next section, where we
present the detailed analysis of the ultraviolet divergences of the
energy-momentum tensor.

For any real $W_k$ and $V_k$ the linear relations (\ref{bogf}) and
(\ref{dbog}) may be inverted to solve for the Bogoliubov parameters
\begin{eqnarray} 
\alpha_{k}(t) &=& i \tilde f_k^*(t)
\left[\dot f_k(t) - \left( i W_k(t) + {V_k(t) \over 2} \right) f_k(t)
\right]\,,\nonumber\\  
\beta_k(t) &=& -i \tilde f_k(t) \left[ \dot f_k(t) + 
\left( i W_k(t)  - {V_k(t) \over 2} \right) f_k(t) \right]\,.
\label{alpbet}
\end{eqnarray}
It is readily verified that these Bogoliubov parameters satisfy the
relation
\begin{eqnarray}
\vert\alpha_k (t) \vert^2 - \vert\beta_k (t) \vert^2 = 1
\; ,
\label{can}
\end{eqnarray}
for each $k$ and for any choice of real $W_k$ and $V_k$, which is what
is required for the transformation to be canonical.  Making use of the
mode equation (\ref{modefns}) and the definition of the adiabatic
functions in (\ref{adbmod}) we obtain the dynamical equations for the
Bogoliubov coefficients
\begin{eqnarray}
\dot\alpha_k &=& -{i\over 2W_k}\left(\Omega_k^2 - W_k^2 
+ {\dot V_k\over 2} + {V_k^2\over 4}\right)(\alpha_k 
+ \beta_k e^{2i\Theta_k})
+ {\beta_k\over 2}e^{2i\Theta_k} \left({\dot W_k\over W_k} 
+ V_k \right)\; , \nonumber\\
\dot\beta_k &=& {i\over 2W_k}\left(\Omega_k^2 - W_k^2 
+ {\dot V_k\over 2} + {V_k^2\over 4}\right)(\alpha_k e^{-2i\Theta_k} 
+ \beta_k)
+ {\alpha_k\over 2} e^{-2i\Theta_k} \left({\dot W_k\over W_k} + V_k \right)\,. 
\label{abdif}
\end{eqnarray}
An equivalent form of the canonical transformation in the Fock space
of creation and destruction operators is given by~\cite{zeldovich}
\begin{eqnarray}
a_{\bf k} &=& \alpha_k^*(t) \tilde a_{\bf k}(t) - \beta^{*}_k(t)
\tilde{a}^{\dagger}_{\bf k}(t)\; , \; \; \; 
{\rm and} \; \; \; 
a_{\bf k}^{\dagger} = \alpha_k(t) \tilde{a}^{\dagger}_{\bf k}(t) - 
\beta_k(t) \tilde a_{\bf k}(t)\,,
\label{boga}
\end{eqnarray}
where the coefficients depend only on the absolute magnitude $k =
|{\bf k}|$ by homogeneity and isotropy.  We may now define the
adiabatic particle number to be
\begin{eqnarray}
{\cal N}_k(t) &\equiv& \langle \tilde a_{\bf k}^{\dagger} (t) \tilde 
a_{\bf k}(t)\rangle
=\vert\alpha_k\vert^2 \langle a^{\dagger}_{\bf k} a_{\bf k}\rangle
+ \vert\beta_k\vert^2 \langle a_{\bf k} a^{\dagger}_{\bf k} \rangle
\nonumber\\  
&=&
\left( 1 + \vert\beta_k\vert^2\right) N_k + \vert\beta_k\vert^2 
\left( 1 + N_k\right)
= N_k + \left( 1 + 2N_k\right)\,\vert\beta_k(t)\vert^2\,.
\label{adbpar}
\end{eqnarray}
If the initial condition on the mode functions is chosen to match the
adiabatic functions, {\it viz.},
\begin{eqnarray}
f_k(t_0) &=& \tilde f_k(t_0)\; , \; \; \; 
{\rm and} \; \; \; 
\dot f_k(t_0)= \left(- i W_k(t) + {V_k(t) \over 2} \right)\tilde f_k(t_0)\,,
\end{eqnarray}
so that 
\begin{eqnarray}
\alpha_k(t_0) &=& 1\;, \; \; \; {\rm and} \; \; \; 
\beta_k(t_0) = 0 \qquad {\rm (adiabatic\ vacuum)}
\; ,
\label{vacinit}
\end{eqnarray}
then the last expression for the adiabatic particle number at
arbitrary time $t$ in (\ref{adbpar}) may be viewed as the sum of the
particle number present at the initial time, $N_k$, plus the particles
created by the time varying geometry, $\vert\beta_k\vert^2$,
multiplied by the Bose enhancement factor, $1 + 2N_k = \sigma_k$, to
account for both spontaneous and induced particle creation processes.

Because the rapidly varying phase $\Theta_k$ drops out of the
definition of the particle number in (\ref{adbpar}), ${\cal N}_k$ is a
relatively slowly varying function of time. In fact, if $W_k$ and
$V_k$ are chosen appropriately to match the adiabatic mode functions
to a given order, ${\cal N}_k$ will be an adiabatic invariant to that
order.  On the other hand the bilinear
\begin{eqnarray}
{\cal C}_k(t) \equiv \langle \tilde a_{\bf k}(t) \tilde a_{\bf k}(t)\rangle
= (1 + 2N_k)\alpha_k\beta^*_k
\; ,
\end{eqnarray}
is a very rapidly varying function of time, since the phase variable
$\Theta_k$ does not drop out of this combination. We can remove the
leading part of this rapid phase variation by defining the quantities
\begin{eqnarray}
{\cal R}_k(t) &\equiv&  {\rm Re}({\cal C}_k e^{-2i \Theta_k}) = 
\sigma_k\,{\rm Re} (\alpha_k \beta_k^*e^{-2i \Theta_k})\,,\nonumber\\
{\cal I}_k(t)&\equiv& {\rm Im}({\cal C}_k e^{-2i \Theta_k}) = 
\sigma_k\,{\rm Im} (\alpha_k \beta_k^* e^{-2i \Theta_k})\,,
\label{RIdef}
\end{eqnarray}
which together with ${\cal N}_k$ span the space of three non-trivial
real bilinears in the adiabatic creation and destruction operators.

Because of the relation (\ref{can}) these three time dependent
functions are not independent, but rather are related by
\begin{eqnarray}
{\cal R}_k^2 + {\cal I}_k^2 &=& \vert{\cal C}_k\vert^2 
= \sigma_k^2 \left(1 + \vert\beta_k\vert^2\right)\,\vert\beta_k\vert^2 
= \left({\cal N}_k + {1\over 2} + {\sigma_k\over 2}\right)
\left({\cal N}_k + {1\over 2} - {\sigma_k\over 2}\right)\nonumber\\
&=&\left({\cal N}_k + {1\over 2}\right)^2 -{\sigma_k^2\over 4}
= \left({\cal N}_k + {1\over 2}\right)^2 - 
\left(N_k + {1\over 2}\right)^2\,,
\label{magC}
\end{eqnarray}
where $\sigma_k$ is a strict constant of motion, dependent only on the
initial data. Differentiating (\ref{adbpar}) with respect to time and
using (\ref{abdif}), and the definitions (\ref{RIdef}), we obtain as
well,
\begin{eqnarray}
{{\mathrm{d}}\over {\mathrm{d}}t} 
{\cal N}_k = {\cal R}_k \left({\dot W_k\over W_k} + V_k \right)
- {{\cal I}_k\over W_k} \left(\Omega_k^2 - W_k^2 
+ {\dot V_k\over 2} + {V_k^2\over 4}\right)\,.
\label{dpart}
\end{eqnarray}
The two relations (\ref{magC}) and (\ref{dpart}) imply that ${\cal
R}_k$ and ${\cal I}_k$ may be eliminated in favor of ${\cal N}_k$ and
$\dot{\cal N}_k$, if desired. Explicitly,
\begin{eqnarray}
{\cal R}_k &=& \vert{\cal C}_k\vert\,\cos (\vartheta_k - \zeta_k)
\; , 
{\cal I}_k = \vert{\cal C}_k\vert\,\sin (\vartheta_k - \zeta_k)\,,
\end{eqnarray}
with the angles $\vartheta_k$ and $\zeta_k$ defined by
\begin{eqnarray}
\vartheta_k &\equiv & 
\sin^{-1} \left[{1\over Z_k}\left({\dot W_k\over W_k} +
V_k \right)\right]\, ,
\nonumber\\ 
\zeta_k &\equiv & \sin^{-1} \left[{\dot{\cal
N}_k\over \vert{\cal C}_k\vert Z_k}\right]\,,
\nonumber\\
Z_k &\equiv & 
\left[\left({\dot W_k\over W_k} + V_k \right)^2 +
{1\over W_k^2} \left(\Omega_k^2 - W_k^2 
+ {\dot V_k\over 2} + {V_k^2\over 4}\right)^2\right]^{1\over 2}\,, 
\end{eqnarray}
and $\vert{\cal C}_k\vert$ is given in terms of ${\cal N}_k$ and the
constant $N_k$ by (\ref{magC}) above. These relations are the general
case of the $SU(1,1)$ structure of the group of Bogoliubov
transformations characteristic of a Gaussian statistical density
matrix, studied previously in the context of the leading order large
$N$ limit of flat space field theory~\cite{diss,phase}.

They will imply in the next section that all components of the
energy-momentum tensor of a free scalar field in a general FRW
background can be expressed in terms of the particle number ${\cal
N}_k$ and its first time derivative, once the adiabatic particle basis
is completely specified by the functions $W_k$ and $V_k$. (This is a
useful fact which suggests an extension of the semiclassical
Boltzmann-Vlasov description of particle creation in cosmological
spacetimes, along the lines of earlier results for particle creation
in time dependent electromagnetic potentials~\cite{kluger}.)

For completeness we give here also the first order differential
equations obeyed by the functions ${\cal R}_k (t)$ and ${\cal I}_k
(t)$,
\begin{eqnarray}
{{\mathrm{d}}\over {\mathrm{d}}t} 
{\cal R}_k &=& \left( {\cal N}_k + {1\over 2}\right)
\left({\dot W_k\over W_k} + V_k \right) + {{\cal I}_k\over W_k}
\left(\Omega_k^2 + W_k^2  + {\dot V_k\over 2} + {V_k^2\over 4}\right)\,,
\nonumber\\
{{\mathrm{d}}\over {\mathrm{d}}t} 
{\cal I}_k &=& -{\left( {\cal N}_k + {1\over 2}\right)\over W_k}
\left(\Omega_k^2 - W_k^2 + {\dot V_k\over 2} + {V_k^2\over 4}\right)
-{{\cal R}_k\over W_k} \left(\Omega_k^2 + W_k^2  + {\dot V_k\over 2} +
{V_k^2\over 4}\right)\,. 
\label{dRI} 
\end{eqnarray}
In all of this discussion we are free to make {\it any} choice of the
time dependent functions $W_k$ and $V_k$ in (\ref{adbmod}),
(\ref{bogf}), and (\ref{dbog}) which is convenient for our purpose,
since no physical quantity can depend on our choice of basis for the
mode functions.  The key point will be to use this freedom to choose
the functions $W_k$ and $V_k$ so that the ultraviolet divergences in
the energy density and pressure may be isolated in the vacuum sector
and removed in a simple way.  This will require that $W_k$ and $V_k$
be matched to the appropriate order of the corresponding adiabatic
functions $W_k^{(n)}$ and $- \dot W_k^{(n)}/W_k^{(n)}$ given for
$n=2$, and $4$ by (\ref{adbtwo}), and (\ref{adbfour}), respectively.

\section{The Energy-Momentum Tensor}

The classical energy-momentum tensor of a free scalar field in an
arbitrary background gravitational field $g_{ab}$ follows from
variation of the action (\ref{action})~\cite{books,Bunch},
\begin{eqnarray}
T_{ab}&=& (\nabla_a \Phi) (\nabla_b \Phi) - \frac{g_{ab}}{2}
g^{cd} (\nabla_c \Phi) (\nabla_d \Phi)
- 2\xi \nabla_a(\Phi\nabla_b \Phi) 
+ 2 \xi g_{ab} g^{cd}\,\nabla_d(\Phi\nabla_c \Phi)\nonumber \\
&+ & \xi G_{ab} \Phi^2  - \frac{m^2}{2} g_{ab} \Phi^2\,,
\end{eqnarray}
and its trace is given by
\begin{eqnarray}
T \equiv g^{ab} T_{ab} =
(6 \xi- 1)g^{ab} \nabla_a (\Phi \nabla_b \Phi) 
- m^2 \Phi^2 -\Phi (-\sq + \xi R + m^2)\Phi\,,
\end{eqnarray}
where the last term vanishes by the equation of motion (\ref{waveq}).

For a perfect fluid in a FRW spacetime we have
\begin{eqnarray}
T_{ab} = p g_{ab} + (\varepsilon + p) v_a v_b\,,
\label{fluid}
\end{eqnarray}
with $v^a$ the components of the velocity field of the fluid.  If the
fluid is comoving with the expansion of the universe, {\it i.e.,} if
it has no peculiar motion with respect to the general expansion, then
$v^a = (1, 0, 0, 0)$ in the coordinates (\ref{FRWs}) and the
energy-momentum tensor is determined completely from the energy
density, $T_{tt} = \varepsilon$ and the trace $T = -\varepsilon + 3
p$. If the quantum state of the scalar field is spatially homogeneous
and isotropic, then $2\langle \Phi \nabla_i \Phi\rangle = \nabla_i
\langle \Phi^2\rangle =0$, and it follows immediately that the
expectation value of its energy-momentum tensor is precisely of the
perfect fluid form (\ref{fluid}).  Hence it suffices to consider
\begin{eqnarray}
\langle T_{tt} \rangle &\stackrel{\rm def}{=}& 
\varepsilon = 
{1 \over 2}\langle \dot\Phi^2 \rangle
-{1\over {2a^2}} \langle \Phi\Delta_3 \Phi \rangle
+ 6 \xi H \langle \Phi \dot \Phi \rangle
+ 3 \xi \left( H^2 + {\kappa \over a^2} \right) \langle \Phi^2 \rangle
+{{m^2} \over 2} \langle \Phi^2 \rangle\,,\nonumber\\
\langle T \rangle &\stackrel{\rm def}{=}&  
g^{ab}\langle T_{ab} \rangle
 = (6\xi-1) \left[-\langle
\dot\Phi^2\rangle  -{1\over a^2}\langle  \Phi\Delta_3\Phi\rangle + 
\xi R\langle \Phi^2 \rangle \right] + 
(6\xi-2) m^2\langle  \Phi^2 \rangle\,,
\label{ttij}
\end{eqnarray} 
with $3p = g^{ij}\langle T_{ij}\rangle$.  Each of the expectation
values of bilinears of the scalar field in this formula are easily
computed as mode sums (or integrals) by inserting the mode expansion
(\ref{modedec}), making use of the definition (\ref{part}), and the
properties (\ref{eigen}) and (\ref{lmsum}) of the harmonic
functions. We find
\begin{eqnarray}
\varepsilon &=& 
\frac{1}{4 \pi^2 a^3} \int \; {\mathrm{d}}
 \mu(k) \; \sigma_k \left[ |\dot f_k|^2
- H {\rm Re}(f_k^* \dot f_k)
+ \left({k^2 \over a^2} + {H^2\over 4} + m^2 \right) |f_k|^2 \right]
\nonumber\\
&+ & \frac{(6 \xi-1) }{2 \pi^2 a^3} \int \; {\mathrm{d}} \mu(k) \; \sigma_k 
\left[H {\rm Re}(f_k^* \dot f_k) + 
\left({\kappa\over 2a^2} - H^2 \right) |f_k|^2\right]\,,
\nonumber \\
\langle T \rangle &=& \frac{m^2}{2 \pi^2 a^3} \int \; {\mathrm{d}} \mu(k) \;
\;  \sigma_k  |f_k|^2 +
\frac{(6 \xi-1) }{2 \pi^2 a^3} \int \; {\mathrm{d}} \mu(k) \; \sigma_k 
\left[ |\dot f_k|^2 +3 H {\rm Re}(f_k^* \dot f_k) + 
\left(\Omega_k^2 + {3\over 2}\dot H\right) |f_k|^2  \right]\,,
\label{epmode}
\end{eqnarray}
with the isotropic pressure given by
\begin{eqnarray}
p = {1\over 3}\left(\varepsilon + \langle T \rangle\right)\,,
\end{eqnarray} 
and
\begin{equation}
\int  {\mathrm{d}} \mu(k) = \left\{ 
\begin{array}{ll} \sum_{k=1}^{\infty} k^2 \; \; & \mbox
{$\kappa = +1$} 
\vspace{0.25cm}
\\ \int_0^{\infty} k^2 {\mathrm{d}} k \; \; &\mbox{$\kappa = 0,\ -1$}
\, .
\end{array}\right. 
\end{equation}

The covariant conservation of the energy-momentum tensor may be
checked by taking the time derivative of $\varepsilon$ and using the
equations of motion (\ref{waveq}) or (\ref{modefns}). By explicit
calculation one can verify that
\begin{eqnarray}
\dot\varepsilon + 3H (\varepsilon + p) = 0\,,
\label{cons}
\end{eqnarray}
provided that in the form (\ref{epmode}) one can take the time
derivative inside the mode sums. Of course this calculation is still
formal since the mode sums actually diverge and require the
introduction of a cutoff or subtractions to become completely
well-defined. In order to be physically meaningful this regularization
and renormalization procedure must preserve the conservation equation
(\ref{cons}).

Now we may use the relations (\ref{adbmod}), (\ref{bogf}), and
(\ref{dbog}), to write the three quantities $\vert f_k\vert^2 $,
$\vert {\dot f}_k \vert^2$, and ${\rm Re}(f_k^* \dot f_k)$ appearing
in the energy density and trace in terms of the three bilinears ${\cal
N}_k$, ${\cal R}_k$, and ${\cal I}_k$ and the two, as yet unspecified,
functions $W_k$ and $V_k$, which define the adiabatic
basis. Explicitly, we have
\begin{eqnarray} 
\sigma_k \; |f_k(t)|^2 
&=& {[1 + 2 {\cal N}_k(t)] \over {2
W_k(t)}} +  {{\cal R}_k(t)\over {W_k (t)}}\,,
\nonumber\\ 
\sigma_k \; |\dot
f_k(t)|^2 &=& 
{[1 + 2 {\cal N}_k(t)] \over {2 W_k(t)}} 
\left( W^2_k(t) + {{V_k^2(t)} \over 4} \right)
+ {{\cal R}_k(t) \over {W_k(t)}} \left[- W^2_k(t) + {V_k^2(t) \over 4} \right]
+ {\cal I}_k(t) V_k(t)\,,
\nonumber\\
\sigma_k \; {\rm Re}[f_k^*(t) \dot f_k(t)]&=&
 {[1 + 2 {\cal N}_k(t)] \over {4  W_k(t)}} V_k(t)
+ {{\cal R}_k(t) \over {2 W_k(t)}}V_k(t) + {\cal I}_k (t)\,.
\label{ffdot}
\end{eqnarray}
Hence we obtain
\begin{eqnarray}
\varepsilon &=&  {1 \over {4 \pi^2 a^3}}\int \; {\mathrm{d}} \mu(k) 
\left[\varepsilon^{\cal N}_{k} \left({\cal N}_k + {1\over 2}\right)
+ \varepsilon^{\cal R}_{k} {\cal R}_k  +  \varepsilon^{\cal I}_{k} {\cal I}_k
\right]\,, \nonumber\\
\langle T \rangle &=&  {1 \over {4 \pi^2 a^3}}\int \; {\mathrm{d}} \mu(k) 
\left[T^{\cal N}_{k} \left({\cal N}_k + {1\over 2}\right)
+ T^{\cal R}_{k} {\cal R}_k  +  T^{\cal I}_{k} {\cal I}_k\right]\,,\nonumber\\
p &=&  {1 \over {4 \pi^2 a^3}}\int \; {\mathrm{d}} \mu(k) 
\left[p^{\cal N}_{k} \left({\cal N}_k + {1\over 2}\right)
+ p^{\cal R}_{k} {\cal R}_k  +  p^{\cal I}_{k} {\cal I}_k\right]\,,
\label{epdecomp}
\end{eqnarray}
where
\begin{eqnarray}
\varepsilon^{\cal N}_k&=& {1 \over W_k} 
\left[W_k^2 + {{V^2_k} \over 4} - {H V_k\over 2} + \omega^2_k + {H^2 \over 4}
+ (6 \xi -1) \left( H V_k + {\kappa \over a^2} - 2 H^2 \right)\right]\,,\nonumber\\
\varepsilon^{\cal R}_k&=& {1 \over W_k} 
\left[- W_k^2 + {V^2_k \over 4} - {H V_k\over 2} + \omega^2_k + {H^2 \over 4}
+ (6 \xi -1) \left( H V_k + {\kappa \over a^2} - 2 H^2
\right)\right]
= \varepsilon^{\cal N}_k - 2W_k \,,\nonumber\\
\varepsilon^{\cal I}_k&=& V_k - H + (6\xi -1) 2 H\,;\\
T^{\cal N}_k&=& {1 \over W_k} \left[-2m^2 + 
2 (6 \xi -1) \left( -W_k^2 - {V^2_k \over 4} + {3 H V_k\over 2} + \omega^2_k
 - {H^2 \over 4} + \dot H\right) + {(6 \xi -1)}^2 {R \over 3}\right]\,,\nonumber\\
T^{\cal R}_k&=& {1 \over W_k} \left[-2m^2 + 
2 (6 \xi -1) \left( W_k^2 - {V^2_k \over 4} + {3 H V_k\over 2} + \omega^2_k
 - {H^2 \over 4} + \dot H\right) + {(6 \xi -1)}^2 {R \over 3}
\right]\nonumber\\
&=&T^{\cal N}_k + 4(6\xi-1)W_k\,,\nonumber\\
T^{\cal I}_k&=& 2(6\xi -1) (3 H -  V_k )\,;\\
p^{\cal N}_k&=& {1 \over 3W_k} \left[W^2_k + {V^2_k \over 4} -  
{H V_k \over 2} + \omega_k^2 -2m^2 + {H^2 \over 4}\right. \nonumber\\
&+ &\left. (6 \xi -1) 
\left( -2 W_k^2 - {V^2_k \over 2} + {4 H V_k} + 2 \omega^2_k
+ 2 \dot H + {\kappa \over a^2} - {5 H^2 \over 2} \right) + 
(6 \xi -1)^2 {R \over 3}\right]\,,\nonumber\\
p^{\cal R}_k&=& {1 \over 3W_k} 
\left[ - W^2_k + {V^2_k \over 4} -  {H V_k \over 2} + \omega_k^2 
-2m^2 + {H^2 \over 4} \right. 
\nonumber\\
&+&\left. 
(6 \xi -1)
\left( 2 W_k^2 - {V^2_k \over 2} + {4 H V_k} + 2 \omega^2_k
+ 2 \dot H + {\kappa \over a^2} - {5 H^2 \over 2} \right)
+ (6 \xi -1)^2 {R \over 3} \right] \nonumber\\
&=& p^{\cal N}_k + 2(4\xi-1)W_k\,,\nonumber\\ 
p^{\cal I}_k&=& {1\over 3}(V_k - H) +{2\over 3}(6\xi -1) (4H - V_k )\,.
\end{eqnarray} 

We are now in a position to determine the specific choice of the
functions $W_k$ and $V_k$, by the requirement that all the power law
divergences of the energy-momentum tensor should be contained in the
vacuum zero point terms, {\em i.e.}, in the terms proportional to the
factors of $1/2$ multiplying $\varepsilon_k^{\cal N}$, $T_k^{\cal N}$,
and $p_k^{\cal N}$, respectively, in Eqn. (\ref{epdecomp}). The
adiabatic expansion of the mode functions $f_k$ may be employed to
isolate these divergences. It has been shown by several
authors~\cite{adiabatic,Bunch,f-p-h} that the fourth order adiabatic
expansion of the frequency $W_k (t)$, if used in (\ref{adbmod}) for
the mode function and then substituted into the expressions for the
energy-momentum tensor above, will reproduce all of the quartic,
quadratic, and logarithmic divergences. The physical reason for this
fact is that for any smoothly varying $a(t)$ the large $k$ behavior of
the exact mode functions is determined by the WKB-like adiabatic form
(\ref{adbmod}), and the local ultraviolet divergences of the
energy-momentum tensor are identical to those obtained in the
adiabatic expansion of the mode functions.  Let us consider the
matching of the large $k$ behavior of the exact and adiabatic mode
functions order by order in the adiabatic expansion.

If we were to take $W_k = W^{(0)}_k = \omega_k$ and $V_k = 0$, then we
would isolate the leading ({\it i.e.,} quartic) divergences in the
energy density and pressure. Indeed, to lowest adiabatic order we
obtain the quartic and quadratic divergent terms,
\begin{eqnarray}
\varepsilon^{(0)}&=& 
{1 \over {4 \pi^2 a^3}}\int \; {\mathrm{d}} \mu(k) \; \omega_k\,,\nonumber\\ 
T^{(0)} &=& - {m^2 \over {4 \pi^2 a^3}}\int \; {\mathrm{d}} \mu(k) 
\; {k^2\over \omega_k}\,.
\label{adbz}
\end{eqnarray}
By differentiating $\varepsilon^{(0)}$ with respect to comoving time,
one can easily show that this adiabatic order zero energy-momentum is
formally conserved. It obeys (\ref{cons}) provided that the time
derivative commutes with the sum (integral) over $k$, {\it i.e.,} that
any cutoff in the mode sum over comoving momentum is time
independent. Thus subtracting it from the full expressions preserves
the conservation law required by general coordinate invariance, even
though the forms (\ref{adbz}) do {\it not} correspond to an explicitly
covariant local counterterm in the effective action.

The reason for this lack of manifest covariance is that the adiabatic
expansion treats time differently than space, and in fact cutting off
the sums in spatial momentum $k$ corresponds precisely to a point
separation in the spacelike hypersurface of constant
$t$~\cite{equivalence,anderson-parker,christensen}. When this is
recognized, then the correspondence between the adiabatic order zero
expressions (\ref{adbz}) and the subtraction of covariant countertems
generated by point splitting in the spacelike direction may be made
fully manifest and becomes completely justified.  In the language of
general renormalization theory, power law divergences are
`non-universal,' in the sense that their precise form depends on the
regularization scheme employed, which in and of itself has no physical
significance. This is clear from the fact that in dimensional
regularization power law divergences do not appear at all. Hence any
convenient scheme, covariant or not, may be employed to isolate and
remove these divergences, although the form of the subtractions will
not be explicitly covariant if the regularization scheme is not, and
the existence of a covariant point splitting procedure equivalent to
the non-covariant adiabatic subtraction is necessary to establish the
latter's validity. The fact that the subtraction of the non-covariant
divergent sums such as (\ref{adbz}) in adiabatic regularization
nevertheless preserves the conservation equation (\ref{cons}) is {\it
a posteriori} evidence enough that the procedure is completely
consistent with general covariance, which is all that we require from
a practical point of view.

The subtraction of the adiabatic order zero expressions from the full
energy density and pressure in (\ref{epdecomp}) leaves subleading
quadratic divergences still present. Hence we must match the function
$W_k$ and $V_k$ to the adiabatic expansion to at least one higher
order. Suppose that we choose to match $W_k = W^{(2)}_k$ to second
adiabatic order. We should allow then for $V_k$ to enter at adiabatic
order one (since $V_k^2$ and $HV_k$, which are then adiabatic order
two, appear in the energy-momentum tensor). Thus to this order we
could choose $V_k =-\dot W_k^{(0)}/W_k^{(0)} =
-\dot\omega_k/\omega_k$.  Indeed this choice will match the quadratic
divergences precisely, and one can verify that the second order
expressions,
\begin{eqnarray} 
\epsilon^{(2)}
&=& 
{1 \over {4 \pi^2 a^3}} \int \; {\mathrm{d}} \mu(k)  
\left\{\omega_k + {H^2 m^4\over 8 \omega_k^5} +{(6 \xi -1)\over 2 \omega_k} 
\left({\kappa \over a^2} - H^2 - {H^2 m^2\over \omega_k^2}\right)\right\}
\,,\nonumber \\
T^{(2)} &=&
 {1 \over {4 \pi^2 a^3}} \int \; {\mathrm{d}} \mu(k)  \left\{ 
-{m^2\over \omega_k} - {m^4\over 4 \omega_k^5} 
(\dot H + 3 H^2) + {5 m^6 H^2\over 8 \omega_k^7} \right. \nonumber\\
&+& \left. {(6 \xi -1)} \left[ {(\dot H + H^2)\over\omega_k}
+ {m^2\over 2 \omega_k^3} \left(2\dot H + 3 H^2 + {\kappa \over a^2} \right)
-{3 m^4 H^2\over 2 \omega_k^5} \right]\right\}\,,
\label{eTtwo}
\end{eqnarray}
obey formal conservation on their own and may be subtracted from the
full expressions in (\ref{epdecomp}). Again the full justification of
this subtraction requires the covariant point splitting analysis. At
this order all the non-universal power law divergences are removed.

Proceeding in this way one more time we might choose $W_k = W^{(4)}_k$
and $V_k = -\dot W_k^{(2)}/W_k^{(2)}$ in order to match the remaining
logarithmic divergences in the vacuum sector and leave behind
completely finite state dependent terms in $\varepsilon$ and $p$. It
is necessary to subtract up to adiabatic order four to obtain the
correct finite trace anomaly in the conformally invariant
limit. However, the remaining logarithmic divergences in the
energy-momentum tensor (unlike the power divergences) correspond
directly to local covariant counterterms in the quantum effective
action, namely the fourth order local invariants, $C_{abcd}C^{abcd}$
and $R^2$. The variation of the first of these vanishes in conformally
flat FRW metrics, and a possible third linear combination of
$R_{ab}R^{ab}$ and $R^2$ vanishes by the invariance of the topological
Gauss-Bonnet density, $R_{abcd}R^{abcd} - 4R_{ab}R^{ab} + R^2$, under
local variations of the metric. Hence only the logarithmic cutoff
dependence of the coefficient of the covariantly conserved tensor
$^{(1)}H_{ab}$, defined by (\ref{Hone}) and (\ref{Honecomp}), remains
after the adiabatic order two terms (\ref{eTtwo}) have been
subtracted.  In fact, it is easy to see that the fourth adiabatic
energy density and trace are given by
\begin{eqnarray}
\label{energy4}
\varepsilon_4&=&\frac{1}{4 \pi^2 a^3} \int {\mathrm{d}} \mu (k)
\frac{1}{4\omega_k^3}
{\left( \xi - \frac{1}{6} \right)}^2 {}^{(1)}H_{tt} 
\nonumber
\\
&+&
\frac{1}{4 \pi^2}
\left\{
\frac{1}{120}(\dot H^2 - 2 H \ddot H - 6 \dot H H^2 -H^4)
+\frac{1}{4}\left(\xi - \frac{1}{6} \right)
\left(6 \dot H H^2 + 2 H \ddot H 
- \dot H^2 -\frac{\kappa}{a^2}H^2  \right)
\right.
\nonumber
\\
&+&
\left.
\frac{3}{2}{\left(\xi - \frac{1}{6} \right)}^2H^2 R \right\}
\; ,
\nonumber
\\
T_4&=&\frac{-1}{4 \pi^2 a^3} \int {\mathrm{d}} \mu (k
\frac{1}{4\omega_k^3}
{\left( \xi - \frac{1}{6} \right)}^2 {}^{(1)}H 
\label{trace4}
\nonumber
\\
&-&
\frac{1}{4 \pi^2}
\left\{
-\frac{7}{840}({\stackrel{\dots}{H}}
+4\dot H^2 + 7 H \ddot H +14 \dot H H^2 +2H^4)
+3\left(\xi - \frac{1}{6} \right)m^2 H^2
\right.
\nonumber
\\
&+&
\left.
\frac{1}{2}\left(\xi - \frac{1}{6} \right)
\left
(\stackrel{\dots}{H} 
 +4\dot H^2 + 7 H \ddot H +12 \dot H H^2 -\frac{\kappa}{a^2} \dot H
-\frac{\kappa}{a^2} H^2 \right)
\right.
\\
\nonumber
&+&
\left.
\frac{9}{2}{\left(\xi - \frac{1}{6} \right)}^2
\left(
\frac{\kappa^2}{a^4}+4 \frac{\kappa}{a^2}\dot H +2  \frac{\kappa}{a^2} H^2
+ 16 H^4 + 30 \dot H H^2 + 4 H \ddot H + 3\dot H^2 \right) \right\}
\; .
\nonumber
\end{eqnarray}
As is evident from these expressions the adiabatic terms of order four
are of two kinds: logarithmically divergent pieces, which are
proportional to the local geometric tensor ${}^{(1)}H_{ab}$, and
finite pieces which are given in terms of the quantities $a, H, \dot
H, \ddot H,$ and $\stackrel{\dots}{H}$. The logarithmic cutoff
dependence of $\langle T_{ab}\rangle$ may be absorbed into a
renormalization of the coefficient of the tensor $^{(1)}H_{ab}$ in the
backreaction equations, {\it i.e.,} by a renormalization of the
coupling constant of the local $R^2$ term in the effective
action~\cite{br}. Since such a term has to be introduced in principle
into the backreaction equations in any case, we may simply include an
explicit logarithmic cutoff dependence (with the correct coefficient)
in the bare $R^2$ coupling to cancel the remaining logarithmic cutoff
dependence in the vacuum energy-momentum tensor of the quantum
field. The resulting backreaction equations will be fully cutoff
independent and obey the covariant conservation relation (\ref{cons}).

In electrodynamic backreaction problems the analogous procedure has
been employed in practical numerical implementations~\cite{largeN}.
The logarithmic divergence in the electric current expectation value
$\langle j^a\rangle$ in that case does not need to be removed by
explicit subtractions. Rather the logarithmic cutoff dependence of the
bare electric charge, $e_0$, is used to cancel the cutoff dependence
of the expectation value of the current $\langle j^a\rangle$ in the
full backreaction equations, $(1/e^{2}_0)\partial_b F^{ab} =
\langle j^a\rangle$.  Provided that the momentum cutoff is chosen much
larger than the inverse scale of temporal variation of the electric
current and fields $F^{ab}$, it can be demonstrated numerically that
the results are independent of the cutoff, as they should be. This can
be checked explicitly by running the numerical code with several
different cutoffs and verifying that the evolution is unchanged,
provided that the bare charge is rescaled logarithmically with the
changed cutoff in such a way as to keep the renormalized charge
fixed. In other words, rather than striving to eliminate it
explicitly, the remaining logarithmic cutoff dependence in the
expectation value of $T_{ab}$ may actually be used to our advantage,
in order to check the cutoff independence in the numerical evolution
of the backreaction equations.

This procedure may be implemented in the semiclassical backreaction
equations as follows.  We propose to avoid explicit subtraction of the
logarithmic cutoff dependence of the energy and pressure, by checking
that it cancels against the logarithmic cutoff dependence of the bare
fourth order coupling multiplying the tensor $^{(1)}H_{ab}$, which
appears in the semiclassical Einstein equation of backreaction. If we
denote the bare coupling of the $R^2$ term in the action by
$1/\alpha_0$ then the logarithmic cutoff dependence in the
backreaction equations will be removed if
\begin{equation}
{1\over \alpha_0} = {1\over \alpha_R (\mu)} + {1\over 16\pi^2}
{\left(\xi - \frac{1}{6} \right)}^2 \log \left({\Lambda\over \mu}\right)\,,
\end{equation}
where $\Lambda$ is the cutoff in the comoving momentum sum (integral)
and $\mu$ is an arbitrary finite renormalization scale.  The finite
adiabatic order four pieces in the energy-momentum tensor are not
taken into account by this procedure, so they can be added back in by
hand to the backreaction equations.

Besides being conceptually and technically somewhat simpler than the
usual approach of explicitly subtracting all terms up to adiabatic
order four, this proposal is a significant simplification for
practical implementation of the backreaction equations on a computer
since $W_k^{(4)}$ as given by (\ref{adbfour}) involves both three and
four time derivatives of the scale factor $a(t)$~\cite{p-s}. These
higher order derivatives present problems for the standard approach to
numerical backreaction calculations, when they occur inside the mode
sums, since one cannot determine the form of $\langle T_{ab}\rangle$
to step the equations forward in time without first knowing these
higher derivatives, which are themselves determined through the
backreaction equations by $\langle T_{ab}\rangle$.

Since in the general FRW backreaction problem we will not remove the
logarithmic cutoff dependence in the energy and pressure connected
with the geometric tensor $^{(1)}H_{ab}$, we will not need to match
the functions $W_k$ and $V_k$ to their full fourth adiabatic order
values. However even if these functions are matched only to second
adiabatic order, fourth order terms in the `vacuum' contributions to
the energy and pressure, $\varepsilon^{\cal N}_k$ and $p^{\cal N}_k$,
will be generated inevitably, from the square of adiabatic order two
terms in $W_k$. In order to cancel these terms in $\varepsilon^{\cal
N}_k$ and $p^{\cal N}_k$ we will choose $W_k$ to be $W_k^{(2)}$ {\it
plus} those additional terms on the first line of (\ref{adbfour})
which fall off only as fast as $k^{-3}$ at large $k$. Likewise we will
choose $V_k$ to be $-\dot\omega_k/\omega_k$ {\it plus} the one
additional term on the first line of (\ref{adbthree}) which falls off
only as fast as $k^{-2}$ at large $k$.

Since we are not attempting to isolate the adiabatic order four
logarithmic divergences, and wish only to avoid introducing spurious
fourth order divergences from the square of $W^{(2)}_k$, we may
evaluate the fourth order terms in $W_k$ and $V_k$ on a spacetime with
{\it vanishing} $^{(1)}H_{ab}$. This is the minimal choice of $W_k$
and $V_k$, which leaves behind only a logarithmic cutoff dependence in
$\varepsilon$ and $p$ proportional to the geometric tensor
$^{(1)}H_{ab}$.  The vanishing of $^{(1)}H_{ab}$, with components
given in eq. (\ref{Honecomp}), allows us to make the following
replacements,
\begin{eqnarray} 
2\dot R H &\rightarrow &R 
\left(\dot H - {\kappa\over
a^2}\right)\,, 
\nonumber\\ 
 \ddot R \rightarrow  -3\dot R
H &\rightarrow &
-{3\over 2}R \left(\dot H - {\kappa\over a^2}\right)\,, 
\end{eqnarray}
when evaluating $W_k$ and $V_k$. Hence the higher derivative terms in 
the first line of (\ref{adbfour}) become
\begin{eqnarray}
\ddot R + 5\dot R H + 2 R\dot H + 6 R H^2 \rightarrow
{R\over 2} \left(R - {8\kappa\over a^2}\right)\,,
\end{eqnarray}
and we finally {\it define}: 
\begin{eqnarray}
W_k &\equiv & 
W_k^{(2)} - {\left(\xi - {1\over 6}\right) 
\over 16\,\omega_k^3}\,R \left(R - {8\kappa\over a^2}\right) - 
{\left(\xi - {1\over 6}\right)^2\over 8\,\omega_k^3}\,R^2
 \nonumber\\
&=&
\omega_k + {\left(\xi - {1\over 6}\right) \over 2\,\omega_k}\,R 
- {m^2\over 4\,\omega_k^3}(\dot H + 3H^2) 
+ {5m^4 \over 8\, \omega_k^5} H^2
\nonumber\\ 
&-& {\left(\xi - {1\over 6}\right) \over 16\,\omega_k^3}\,R 
\left(R - {8\kappa\over a^2}\right) - 
{\left(\xi - {1\over 6}\right)^2\over 8\,\omega_k^3}\,R^2 \,.
\label{Wdef}\\
V_k &\equiv & -{\dot\omega_k\over \omega_k} - 
{\left(\xi - {1\over 6}\right) \over \omega_k^2}\,RH -
{\left(\xi - {1\over 6}\right) \over 4\,\omega_k^2}\, 
{R\over H} \left(\dot H - {\kappa \over a^2}\right)\nonumber\\
&=& H\left(1 - {m^2\over\omega_k^2}\right) - 
{\left(\xi - {1\over 6}\right) \over \omega_k^2}\,RH -
{\left(\xi - {1\over 6}\right) \over 4\,\omega_k^2}\, 
{R\over H} \left(\dot H - {\kappa \over a^2}\right)
\; ,
\label{Vdef}
\end{eqnarray}
to specify fully the adiabatic particle number basis for the scalar
field of arbitrary mass and curvature coupling in a general FRW
spacetime.

Some comments about this definition and the corresponding choice of
basis bear emphasizing. First, by construction these choices for $W_k$
and $V_k$ match all the quartic and quadratic power law divergences in
the energy and pressure in a general FRW spacetime. Thus the
subtraction of the adiabatic two expressions (\ref{eTtwo}) from the
bare energy density and trace will leave behind only the logarithmic
cutoff dependence proportional to $^{(1)}H_{ab}$, whenever it is
non-vanishing.

Second, with these requirements, the choice of $W_k$ and $V_k$, and
hence the adiabatic basis they define is the almost unique, minimal
choice. Since the adiabatic (order zero and two) matching is required
to eliminate the non-universal power divergences in the
energy-momentum tensor, and since the matching of any higher adiabatic
orders requires higher order time derivatives of the scale factor
which we exclude, the only freedom left in the choice of $W_k$ and
$V_k$ are adiabatic order four terms in $W_k$ and third order terms in
$V_k$, which either give rise to local $^{(1)}H_{ab}$ terms in
$\langle T_{ab}\rangle$ or fall off faster than $1/k^3$ and hence lead
to finite reapportionment of the terms in the energy-momentum tensor
between `particle' and `vacuum' contributions. This residual ambiguity
in the separation of the energy-momentum terms into particle and
vacuum contributions cannot be removed except by an essentially
arbitrary choice. It reflects the necessary uncertainty in the
definition of the `particle' concept in time varying external
fields. However the residual ambiguity is fairly mild and does not
affect severely the physical interpretation of particles in
semiclassical cosmology where the variations of the metric are assumed
small in Planck units. Our definitions of $W_k$ and $V_k$ in
(\ref{Wdef}) and (\ref{Vdef}) are the {\it minimal} ones, involving no
more than second derivatives of the metric, that make such an
interpretation of the separation of $\langle T_{ab}\rangle$ into
quantum vacuum and quasiclassical particle components possible for
slowly varying $a(t)$.

Third, for the massless, conformally coupled scalar, $m=0$ and $\xi =
{1\over 6}$, the definitions (\ref{Wdef}) and (\ref{Vdef}) simplify
considerably: $W_k = k/a$ and $V_k = H$. Referring back to
(\ref{bigom}) and (\ref{dpart}) we observe that the adiabatic particle
number ${\cal N}_k$ is {\it constant} in this case, {\it i.e.,} ${\cal
N}_k =N_k$. Hence there is no creation of massless, conformally
coupled field quanta in arbitrary FRW spacetimes, as expected from
conformal invariance.

Fourth, for the massless, minimally coupled scalar field, $m=\xi=0$,
\begin{eqnarray}
W_k &=& {k\over a}\left( 1 - {Ra^2\over 12 k^2} \right)^2 + {Ra\over 12 k}
\left( 1 - {\kappa \over k^2}\right)\,,\nonumber\\
V_k &=& H + {RHa^2\over 6 k^2} + {Ra^2\over 24 k^2 H} \left( \dot H -
{\kappa\over a^2}\right)\,, \; \; \; {\rm for} \; \;\ \; m=\xi =0 
\; .
\label{mmc}
\end{eqnarray}
We observe that $W_k$ is everywhere non-negative for $R \ge 0$,
vanishing only for the spatially constant mode in the case of closed
spatial sections, $k=\kappa = 1$, and then only at times for which $R
= 12/a^2$.  With these special exclusions the definition of adiabatic
particle basis specified by (\ref{Wdef}) and (\ref{Vdef}) is
meaningful even in this quite infrared sensitive case, which is the
most analogous to linearized graviton fluctuations.

In this paper we restrict ourselves to a test field approximation, and
do not consider the backreaction of the created `particles' in the
background geometry. For the special case of de Sitter, which will be
the particular spacetime under study, the tensor $^{(1)}H_{ab}$
vanishes.  Hence in this fixed background the one remaining
logarithmic cutoff dependence proportional to the geometric tensor
$^{(1)}H_{ab}$ (after subtraction of the adiabatic order two
expressions) will vanish identically, and the resulting renormalized
energy-momentum tensor is fully cutoff independent, without any
additional countertems. In order to compare the value of the finite
$\langle T_{ab} \rangle$ with previous authors, we perform the final
adiabatic subtraction of order four as well. The fact that
${}^{(1)}H_{ab}$ is zero for de Sitter simplifies this order four
subtraction, so that we only need to subtract the finite pieces of
(\ref{energy4}) and (\ref{trace4}), and the logarithmic cutoff
dependent counterterm is not needed to obtain the renormalized
energy-momentum tensor.

Examining in detail the adiabatic `vacuum' component of the energy and
pressure, obtained by subtraction of the second order adiabatic
expressions (\ref{eTtwo}),
\begin{eqnarray} 
\varepsilon_V &\equiv &
 {1 \over{8\pi^2 a^3}} \int \; {\mathrm{d}} 
\mu(k)  \left( \varepsilon_k^{\cal N} 
- \varepsilon^{(2)} \right)
= {1 \over
{8\pi^2 a^3}} \int \; {\mathrm{d}} \mu(k)  \bar\varepsilon_k
\; ,
 \nonumber\\ 
p_V &\equiv & 
{1 \over {24\pi^2 a^3}} \int \; {\mathrm{d}} \mu(k)  \left(\varepsilon_k^{\cal
N}+ T_k^{\cal N}\right) - {1\over 3}\left(\varepsilon^{(2)} +
T^{(2)}\right) 
= {1 \over {24\pi^2 a^3}} \int \; {\mathrm{d}}
\mu(k)  \left(\bar\varepsilon_k +
\bar T_k\right) 
\; ,
\label{vacep}
\end{eqnarray}
with
\begin{eqnarray}
\bar\varepsilon_k &=&
 {1\over W_k}\left[W_k -\omega_k - 
{\left(\xi -{1\over 6}\right)\over 2\omega_k}R\right]
\left[W_k -\omega_k - 
{\left(\xi -{1\over 6}\right)\over 2\,\omega_k}R - 
{(6\xi -1)\over \omega_k}\left(HV_k - 2H^2 + {\kappa\over a^2}\right)\right]
\nonumber\\
&+& {(V_k - H)^2\over 4\,W_k} - {3\,\left(\xi -{1\over 6}\right)^2
(V_k - H)\over\omega_k^2\,W_k}\,RH - {3\left(\xi -{1\over 6}\right)^2
(W_k - \omega_k)\over 2\,\omega_k^4\,W_k}\,R
\left(\dot H + 4H^2 + {\kappa\over 2a^2}\right)\,,
\end{eqnarray}
and
\begin{eqnarray}
\bar T_k &=& 
{2\over W_k}\left[W_k -\omega_k - 
{\left(\xi -{1\over 6}\right)\over 2\omega_k}R\right]
\left[{m^2\over\omega_k} + (6\xi -1)(W_k -\omega_k) 
- {3\left(\xi -{1\over 6}\right)^2\over \omega_k}R 
- {(6\xi -1)\over\omega_k}(\dot H + H^2)\right]
\nonumber\\
&-& {(6\xi -1)(V_k - H)\over 2\,W_k}\left[V_k - H 
+ {4 H\over \omega_k}(W_k - \omega_k)\right]
+ {5 m^4\over 4 \omega_k^5}\left(1 - {m^2\over \omega_k^2}\right)H^2
\nonumber\\
&+& {\left(\xi -{1\over 6}\right) 
(W_k - \omega_k)\over \omega_k^3W_k}R
\left[-m^2 + 3\left(\xi -{1\over 6}\right)R 
+ (6\xi -1)(\dot H + H^2)\right]\,,
\end{eqnarray}
and using the definitions of $W_k$ and $V_k$ in (\ref{Wdef}) and
(\ref{Vdef}), we observe that the sums (integrals) for the these
vacuum contributions to the energy density and pressure in
(\ref{vacep}) are explicitly convergent for arbitrary mass and
curvature coupling $\xi$ in a general FRW spacetime. Hence the
remaining logarithmic cutoff dependence of $\langle T_{ab}\rangle$
proportional to $^{(1)}H_{ab}$ must reside only in the terms linear in
${\cal N}_k$, ${\cal R}_k$, and ${\cal I}_k$ in (\ref{epdecomp}), and
this remaining logarithmic cutoff dependence vanishes as well in the
special case of de Sitter spacetime.

An additional technical point is the fact that when performing the
adiabatic subtraction one should use the continuous measure, and
not the discrete one, even in the case of a closed cosmological
model~\cite{anderson-parker}. This scheme is necessary to compute
finite Casimir energies and the correct trace anomaly, for example.
We only point out at this
stage the relevant steps needed to carry out this subtraction. The
details of this calculation can be found in~\cite{anderson-parker}.

The terms needed are called the Plana terms, and are given by
\begin{eqnarray}
{\cal P}_{-n}&\stackrel{\rm def}{=}&\sum_{k=1}^{+\infty} k^2 \; k^{-n}   - 
\int_{0}^{+\infty} {\mathrm{d}} k \; k^2 \; k^{-n}   
=
-\int_{0}^{1} {\mathrm{d}} k 
\; k^2 \; k^{-n}   + \frac{1}{2}-2 \int_{0}^{+\infty}
{\mathrm{d}} z \frac{q(z)}{e^{2\pi z} - 1}
\; ,
\end{eqnarray}
with
\begin{eqnarray}
q(z)=\frac{1}{2i} \left[ \frac{{(1+iz)}^2}{{(1+iz)}^n}
-\frac{{(1-iz)}^2}{{(1-iz)}^n}
\right]
\; .
\end{eqnarray}
We need to calculate the Plana terms for $n=-1$, $n=1$, and $n=3$, as
these values of $n$ correspond to the divergences present in the
theory. It is easy to see that the calculation of the Plana term for
$n=3$ requires the introduction of an infrared regulator $\zeta$. With
this is mind, we have
\begin{eqnarray}
{\cal P}_1=\frac{1}{120}\; , \; \; \; {\cal P}_{-1}=\frac{-1}{12}\; , \; \; \; 
{\cal P}_{-3}=\log \zeta + \gamma_{E}=
\log \zeta
- \psi(1)=\log \zeta + 0.5772157 \dots
\; .
\end{eqnarray}
The infrared regulator drops out of the final answer for the
renormalized $\langle T_{ab} \rangle$ when the state is IR finite. We
present results for the renormalized $\langle T_{ab} \rangle$ for the
special case of a de Sitter universe in the next section, with these
additional finite Plana terms subtracted (as described
in~\cite{anderson-parker}) in order to compare our results to the
earlier literature.

\section{Particle Creation in de Sitter Spacetime}

The line element of de Sitter space with $\kappa=+1$ is given by
(\ref{FRWs}) and (\ref{spatial}) with
\begin{equation}
a_{deS}(t)=\bar H^{-1}\cosh \bar Ht\, .
\label{desitter}
\end{equation}
Because of its higher degree of symmetry and special status both in
the preceeding discussion of renormalization of $\langle
T_{ab}\rangle$ and in cosmological models of the early universe, we
will apply our decomposition of the renormalized energy-momentum
tensor of the scalar field first to de Sitter spacetime. The de Sitter
invariant `vacuum' state (or variants thereof) is by far the most
commonly discussed in the literature~\cite{allen}. However, from the
point of view of the initial value problem for the scalar field
theory, specified on a complete spacelike Cauchy surface, there is no
{\it a priori} reason to impose the global $SO(4,1)$ invariance of de
Sitter spacetime. Our general framework permits us to consider {\it
any} spatially homogeneous and isotropic initial state consistent with
finite, renormalized energy and pressure.

This general initial state of the scalar field is specified by initial
data on the mode functions, $f_k (t_0)$ and $\dot f_k(t_0)$,
constrained to obey the Wronskian condition (\ref{wron}), together
with the initial particle number density in phase space ${\cal
N}_k(t_0)$. By convention and without any loss of generality we will
retain our previous definition of $N_k$ to be the initial particle
number density with respect to the initial adiabatic mode basis
specified by (\ref{adbmod}), (\ref{Wdef}), and (\ref{Vdef}). This
means that for general initial data we must allow $f_k (t_0)$ and
$\dot f_k(t_0)$ to differ from their adiabatic vacuum values $\tilde
f_k (t_0)$ and $\dot {\tilde f_k}(t_0)$, respectively. Consequently,
the Bogoliubov coefficients $\alpha_k (t_0)$ and $\beta_k (t_0)$ do
not satisfy (\ref{vacinit}) and ${\cal N}_k (t_0) \ne N_k$ in general.

The global phase freedom in the mode functions allows us to choose
$f_k(t_0)$ real. Thus we may define the two real functions of $k$,
$w_k$ and $v_k$, by
\begin{eqnarray}
f_k(t_0) &\equiv & {1\over \sqrt{2 w_k}}\,,\nonumber\\
\dot f_k(t_0) &\equiv & \left(-iw_k + {v_k\over 2}\right)f_k(t_0)\,,
\label{initcon}
\end{eqnarray}
which is the general form of initial data satisfying the Wronkskian
condition (\ref{wron}). The physical condition that the initial state
have finite energy and pressure requires that $w_k$ and $v_k$ must
match the adiabatic vacuum values $W_k$ and $V_k$, defined by (\ref{Wdef})
and (\ref{Vdef}), up to terms that fall off sufficiently fast at 
large $k$, {\it i.e.},
\begin{eqnarray}
\vert w_k - W_k \vert &<  &{\cal O}\left(k^{-3}\right)\,,\nonumber\\
\vert v_k - V_k \vert &<  &{\cal O}\left(k^{-2}\right)
\; ,
\label{wvasy}
\end{eqnarray}
as $k \rightarrow \infty$. Likewise we must require that the initial
particle distribution have finite energy and pressure, so that
\begin{eqnarray}
N_k < {\cal O}\left(k^{-3}\right)\qquad {\rm as} 
\quad k \rightarrow \infty\,.
\label{Nasy}
\end{eqnarray}
Any choice of finite initial data in the form of the three functions
$w_k$, $v_k$, and $N_k$ satisfying these conditions at large $k$ have
finite initial renormalized energy and pressure and are physically
allowed.  This condition of finite initial energy and pressure is
sufficient to guarantee finite energy and pressure for the scalar
field at all subsequent times.

In the special case of spacetimes for which $^{(1)}H_{ab} =0$, such as
de Sitter space, we may saturate the inequalities (\ref{wvasy}) and
(\ref{Nasy}), since the logarithmic ultraviolet divergences in
$\langle T_{ab}\rangle$ this would lead to in a general FRW space are
proportional to $^{(1)}H_{ab}$ and vanish when $^{(1)}H_{ab}$
vanishes.

The numerical solution of the initial value problem proceeds as
follows. We first define initial conditions as given in
eq. (\ref{initcon}), and then solve the mode equations using a sixth
order Runge-Kutta integrator.  All quantities of interest can be
derived from the numerically evaluated $f_k(t)$ and $\dot{f}_k(t)$.
Computations of the energy-momentum tensor involve direct summations
(not integrals) since we are considering closed spatial sections.
Since all the modes are independent the calculation is well-suited to
a parallel computer.

\begin{figure}
\epsfxsize=15cm
\epsfysize=8cm
\centerline{\epsfbox{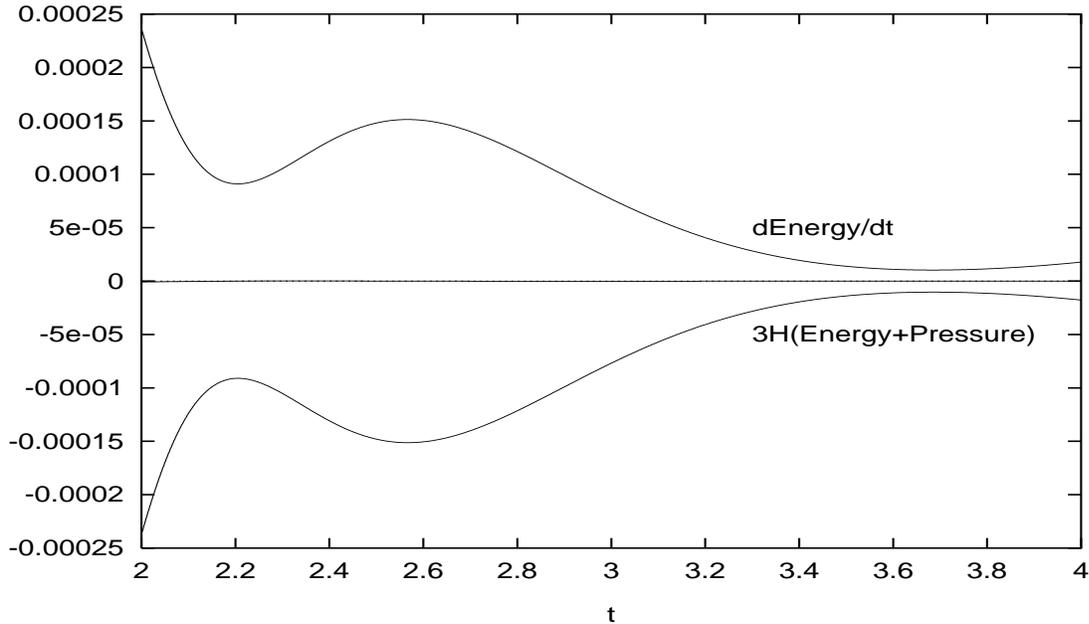}}
\vspace{0.25cm}
\caption
{The top line plots $\frac{d \varepsilon_{\rm {\small R}}}{dt}$, the
bottom line plots $3 H (\varepsilon_{\rm {\small R}}+ p_{\rm {\small
R}})$, and the sum of both is vanishing, showing that the renormalized
energy-momentum tensor is covariantly conserved.}
\label{conservation}
\end{figure}

The numerical scheme preserves the covariant conservation of both the
bare and the renormalized $\langle T_{ab} \rangle$. The conservation
of the bare tensor is a direct result of the equation of motion for
the field $\Phi(t,{\bf x})$.  Because the adiabatic subtraction
procedure at any given order preserves the covariant conservation of
the renormalized $\langle T_{ab} \rangle$, $\langle T_{ab} \rangle_R$
must be conserved as well.  In Fig. \ref{conservation} we show
explicitly the covariant conservation of the renormalized
energy-momentum tensor for $m=1.0$, $\xi=1/6$, $t_0=2.0$, and a
momentum cutoff of $\Lambda=400$.
One can also show explicitly that our results are cutoff independent.
That is, if we run with a momentum cutoff $\Lambda$ (the total number
of modes is $\Lambda$), all renormalized quantities remain cutoff
independent until we reach the time at which all the modes have
crossed the horizon. That is, we can trust our results until times of
order of $t_\Lambda$, with $t_\Lambda$ given by $\cosh (\bar H
t_\Lambda)
\approx \Lambda$. 

\begin{figure}
\epsfxsize=15cm
\epsfysize=8cm
\centerline{\epsfbox{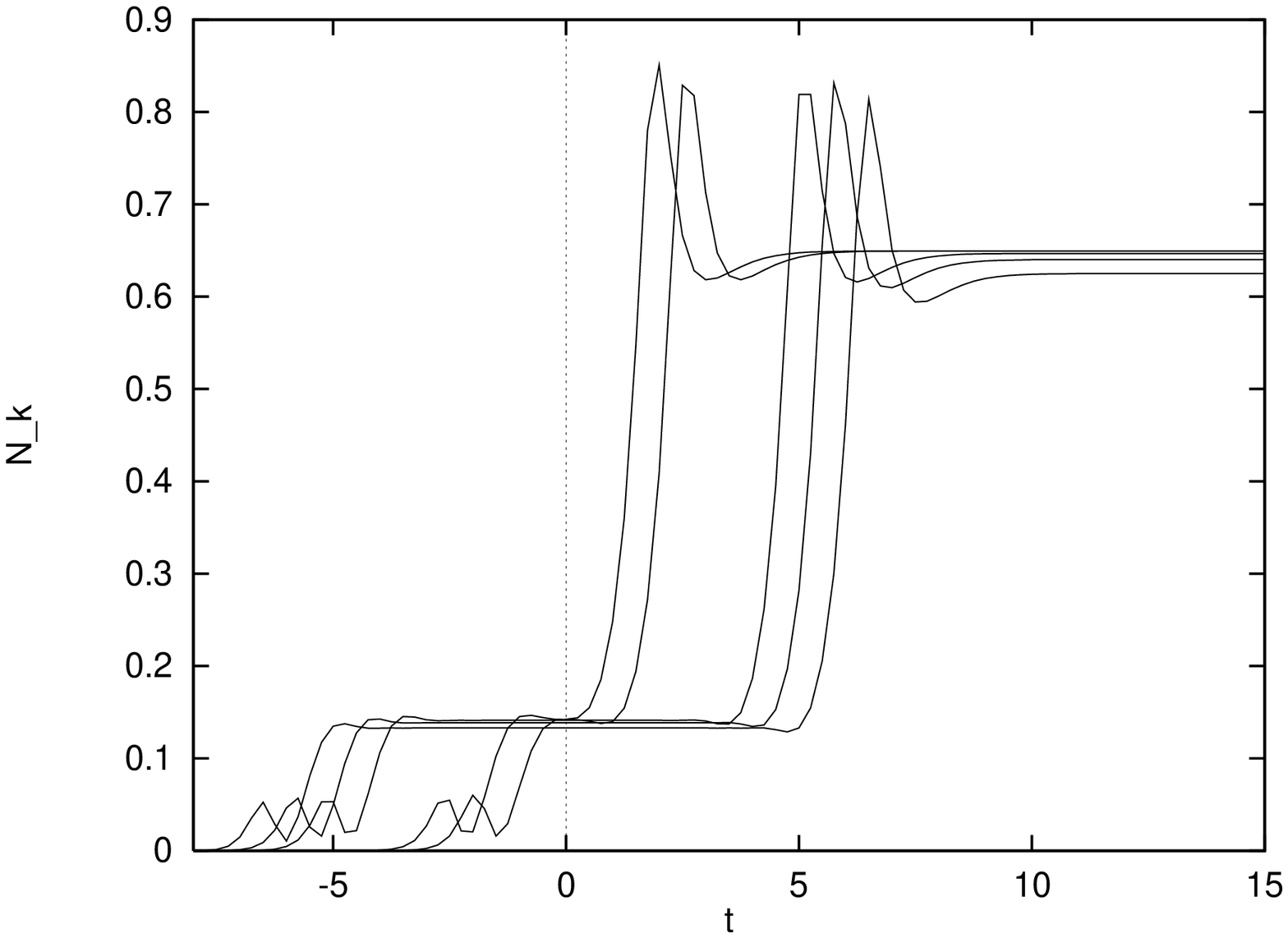}}
\vspace{0.25cm}
\caption
{Particle number for $m=0.6$, $\xi=1/6$, $t_0=-8$, and different
values of comoving momentum, from left to right $k=1,2,10,20,25,50,100$.}
\label{particle-8}
\epsfxsize=15cm
\epsfysize=8cm
\centerline{\epsfbox{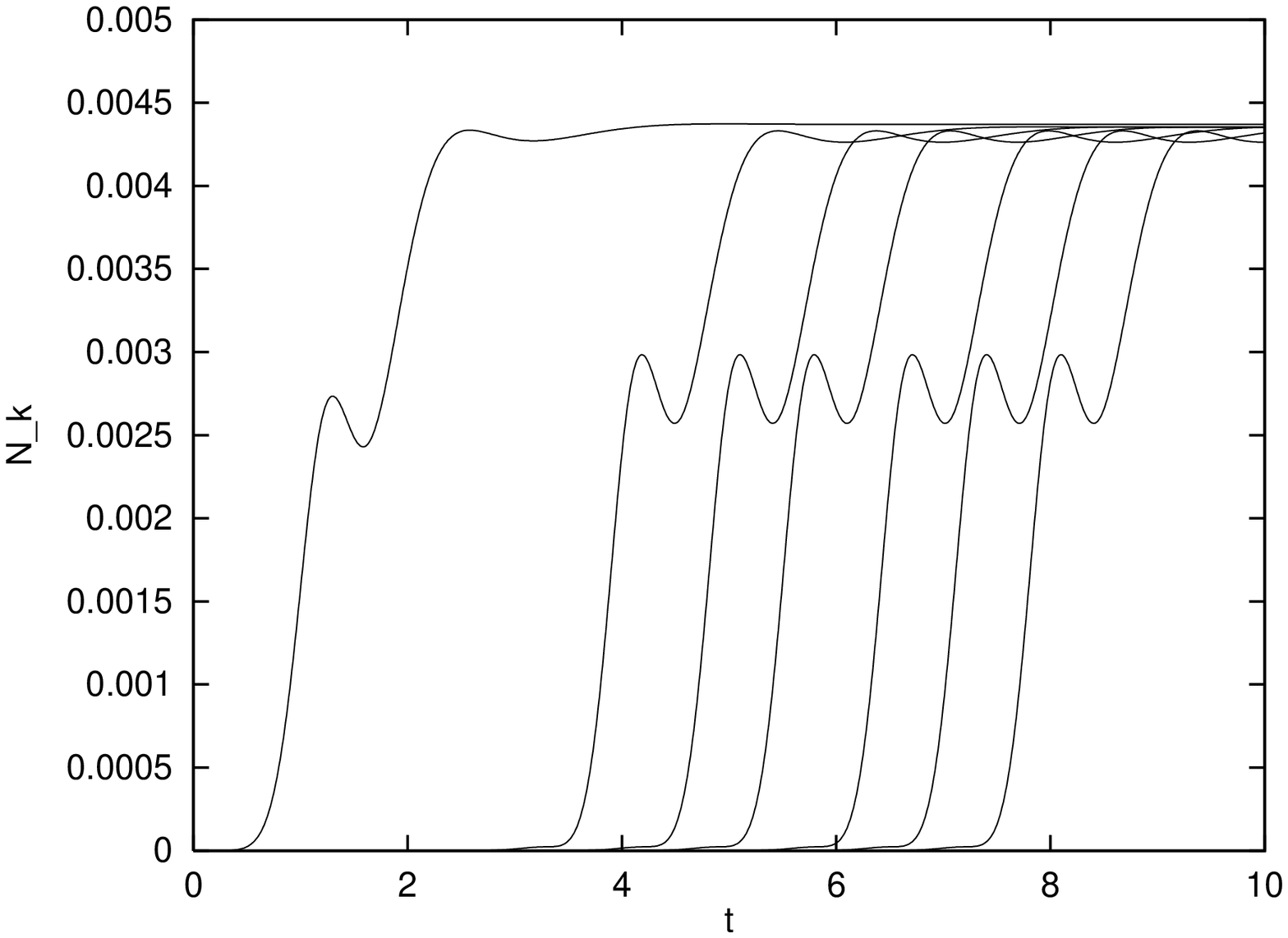}}
\caption
{Particle number for $m$=1.0, $\xi$=1/6, $t_0=0$, and different values
of comoving momentum, from left to right $k$=1, 20, 50, 100, 200, 250,
500, 1000.}
\label{particle6}
\epsfxsize=15cm
\epsfysize=8cm
\centerline{\epsfbox{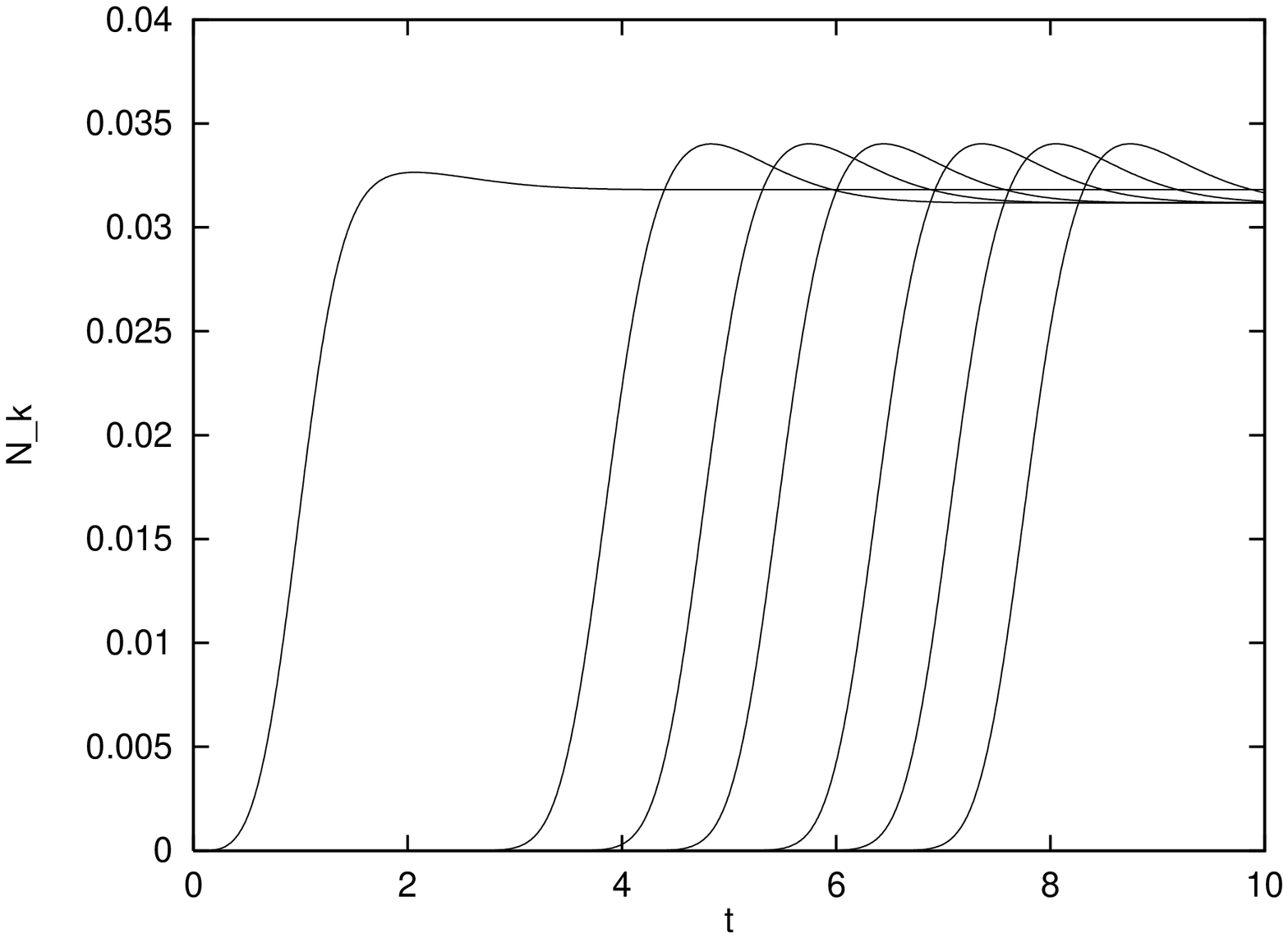}}
\caption{Particle number for $m$=1.6, $\xi$=0, $t_0=0$, and 
different values of comoving momentum,
from left to right,
$k$=1, 20, 50, 100, 200, 250, 500, 1000.}
\label{particle0}
\end{figure}

\begin{figure}
\epsfxsize=15cm
\epsfysize=8cm
\centerline{\epsfbox{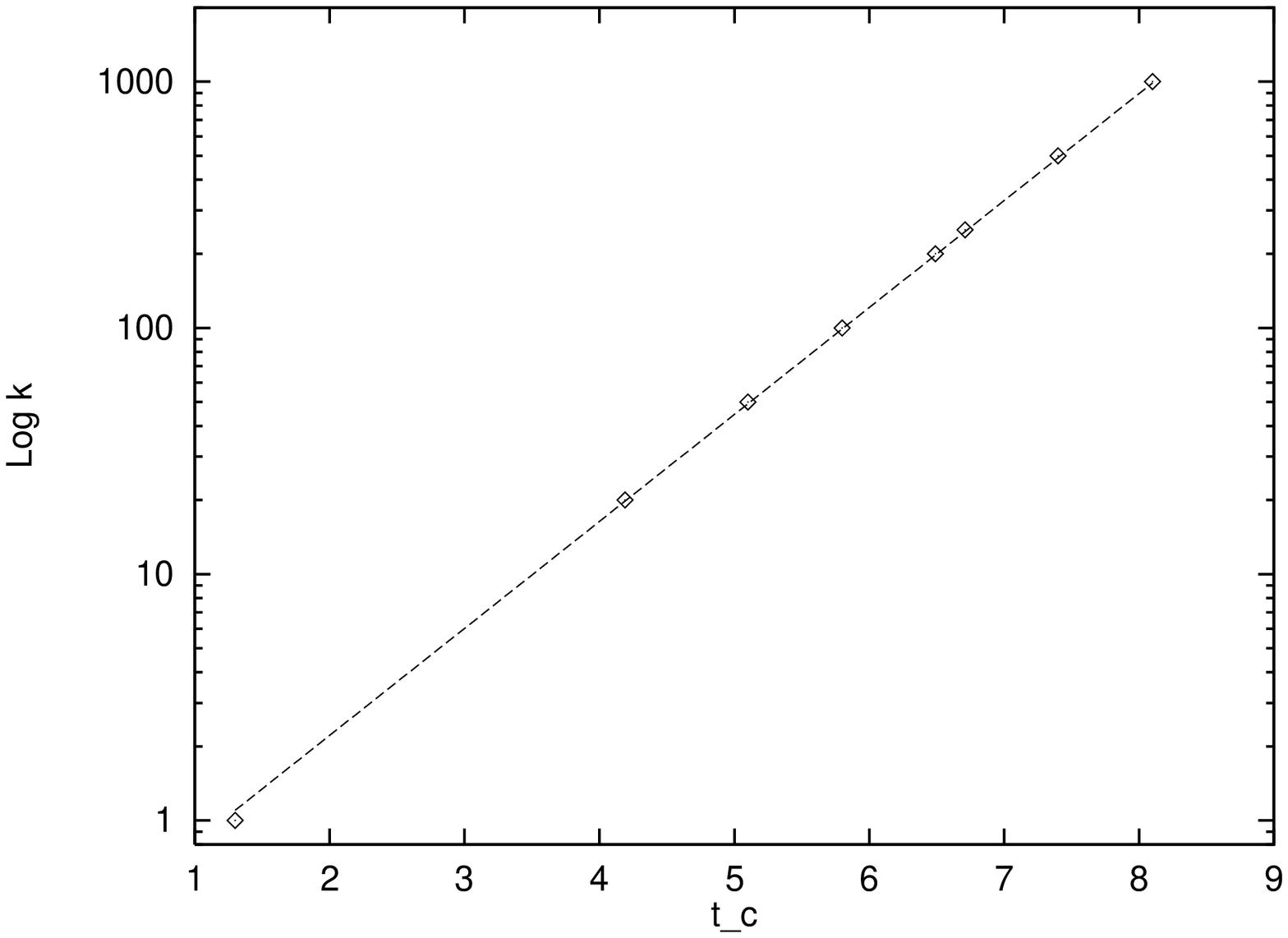}}
\caption
{Momentum versus comoving crossing time for $m$=1.0, $\xi$=1/6, and 
$k$=1, 20, 50, 100, 200, 250, 500, 1000.}
\label{horizon6}
\epsfxsize=15cm
\epsfysize=8cm
\centerline{\epsfbox{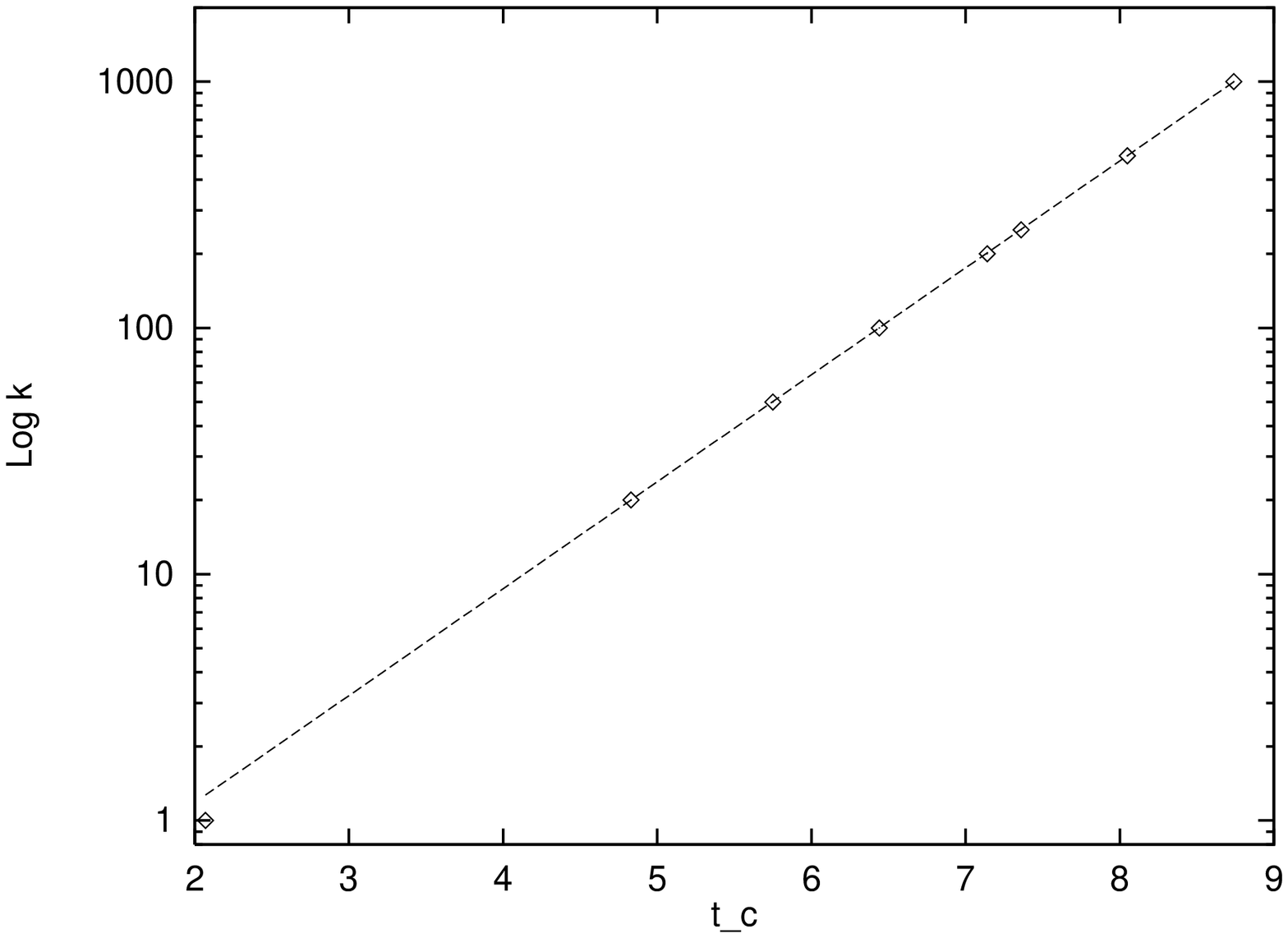}}
\caption
{Momentum versus comoving crossing time for $m=1.6$, $\xi=0$, and
$k$=1, 20, 50, 100, 200, 250, 500, 1000.}
\label{horizon0}
\end{figure}

In Figs. 2 and 3 we plot the particle number ${\cal N}_k (t)$ in de
Sitter space for various values of $k$, $m$, and $\xi$, starting at
$t_0 =-8$ and $t_0=0$ (in units of $\bar H = R/12$) with adiabatic
vacuum initial conditions, and the choices $w_k = W_k$ and
$v_k=V_k$. When $m$ and $\xi$ are such that
\begin{eqnarray}
\gamma^2\equiv {M^2 \over \bar H^2} - {9\over 4} = {m^2\over \bar H^2} +
12\xi - {9\over 4} > 0\,, \end{eqnarray} the particle number increases
sharply (with a rise time of order $\bar H^{-1}$) when the physical
wavelength of the mode crosses the de Sitter horizon, {\it i.e.,} when
$t = t_c$, where
\begin{eqnarray}
\cosh(\bar H t_c) \simeq k\,.
\label{horiz}
\end{eqnarray}
When the initial time $t_0 \ge 0$, the de Sitter universe (in closed
spatial coordinates) is always expanding thereafter and each mode can
cross the horizon and go through its particle creation event at most
once. This situation is shown in Fig. \ref{particle6} and
Fig. \ref{particle0}.  If the initial time $t_0 \le 0$, the spatial
sections first contract, then expand, and some modes can cross the
horizon twice, once on the way in and then again on the way out,
corresponding to the two solutions of (\ref{horiz}) at $t=\pm t_c$.
These modes then go through two separate bursts of particle creation,
as illustrated in Fig. \ref{particle-8}. The behavior of the horizon
crossing time with $k$ in accordance with (\ref{horiz}) is shown in
Fig. \ref{horizon6} and Fig. \ref{horizon0}.
In Figs. 5 and 6 we plot the logarithms of the comoving momentum as a
function of their crossing time $t_{\rm c}$. The lines correspond to
the fits, $0.3 \exp t$ and $0.16 \exp t$, for $\xi=1/6$ and $\xi=0$,
respectively, in agreement with the previous considerations regarding
the nature of the particle production at horizon crossing and earlier
work on stochastic inflation~\cite{starobinsky,habib}.
This corresponds to the physical picture of particle creation as a
parametric resonance between the wave function of the particle mode
and the geometry, which is maximized when the physical wavelength of
the mode is the same as the horizon scale.

In Figs. 7 and 8 we have plotted the renormalized values of the energy,
trace, and pressure for the cases considered above.
\begin{figure}
\epsfxsize=15cm
\epsfysize=8cm
\centerline{\epsfbox{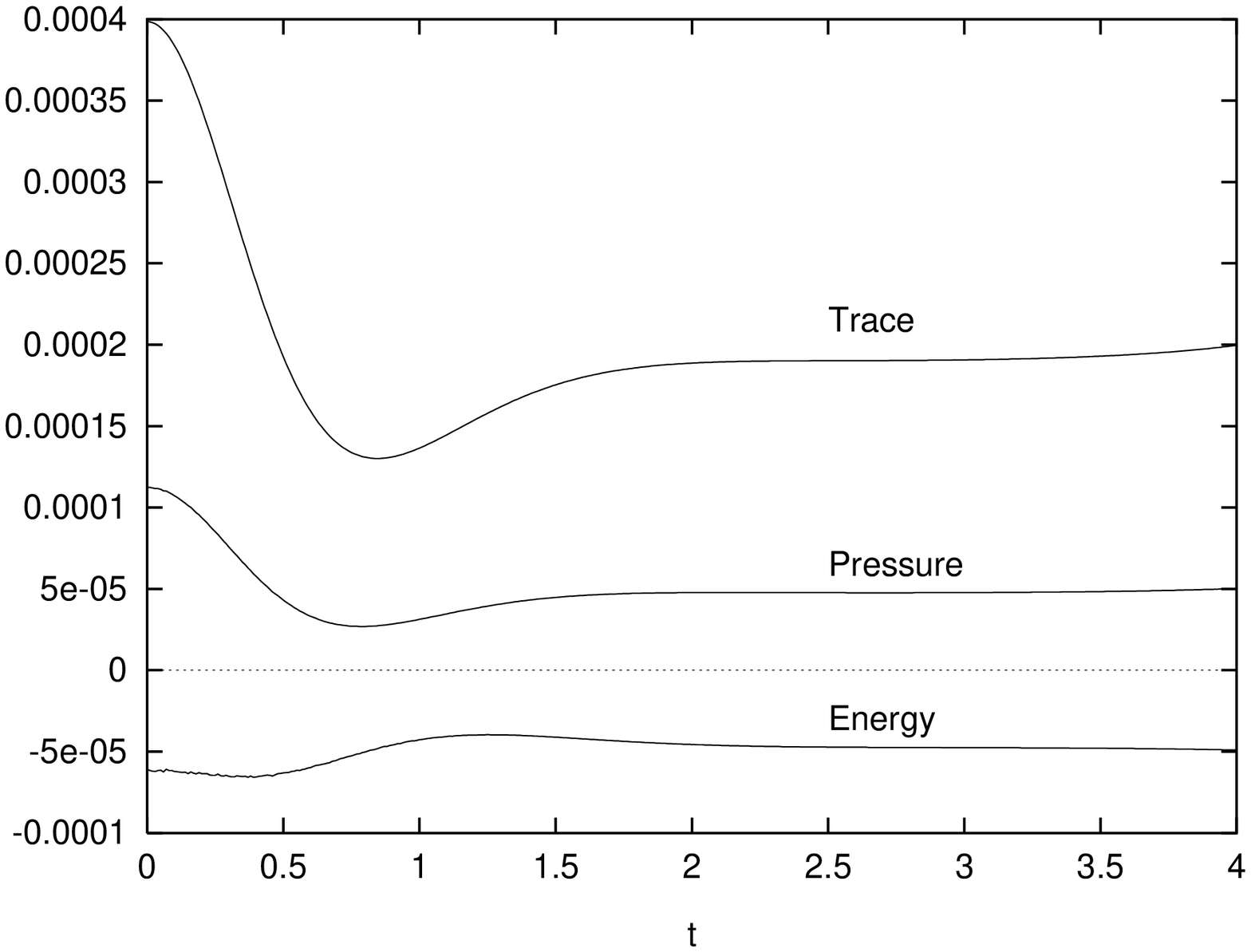}}
\vspace{0.25cm}
\caption
{Energy density, trace, and pressure for $t_0=0$, $m=1.0$, and $\xi=1/6$.}
\label{etp-6}
\epsfxsize=15cm
\epsfysize=8cm
\centerline{\epsfbox{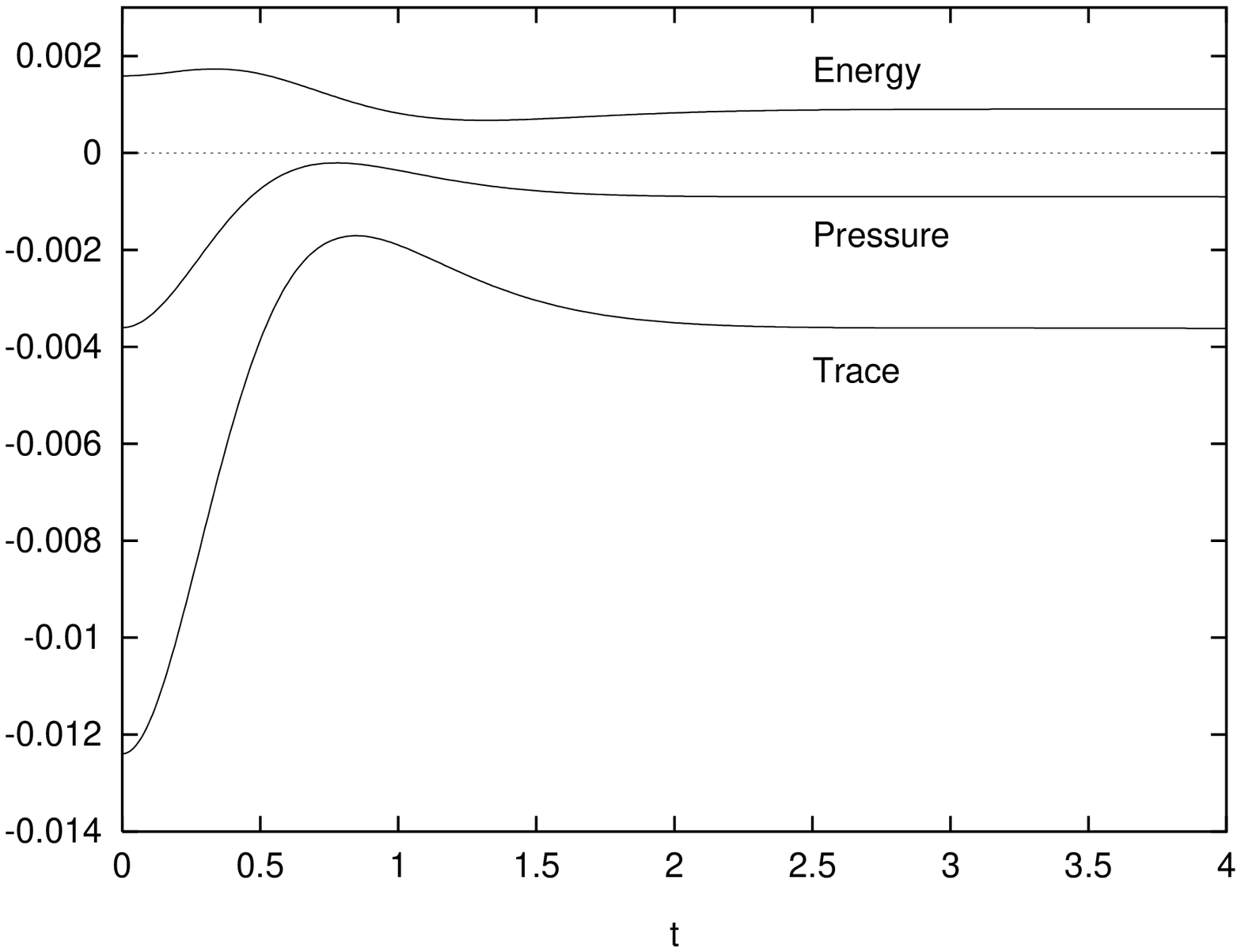}}
\vspace{0.25cm}
\caption
{Energy density, trace, and pressure for $t_0=0$, $m$=1.6, and $\xi$=0.}
\label{etp-0}
\end{figure}
The most important feature of the figures to note is that the particle
number goes to a constant which is almost independent of $k$ at late
times, when the modes freezes out. The value of the constant plateau
value of ${\cal N}_k$ at late times and large $k$ can be compared with
that calculated by one of us~\cite{Mottola} analytically some years
ago, namely ${\cal N}_k \rightarrow
\sinh^{-2}(\pi\gamma)$, assuming an initial `in' vacuum state
corresponding to pure positive frequency as $t_0\rightarrow
-\infty$. Although this `in' state is a higher order adiabatic vacuum
state than the one considered in the present work, the difference
between the two goes to zero rapidly for large $\gamma$.
We remark that if we had chosen a
different separation of particle and vacuum components
of $\langle T_{ab}\rangle$ by making a slightly different definition of
$W_k$ and $V_k$ at adiabatic order four, then the detailed
time profile of the ${\cal N}_k$ near the horizon crossing would
be slightly different, but the location of the crossing time
and the plateau values of particle number at later times observed
in the figures would not be altered.

The fact that the particle number density in any given $k$ mode does
not redshift away despite the exponential expansion is related to the
fact that these created particles do not satisfy a standard dust or
matter equation of state with $\varepsilon \ge 0$, $p \ge 0$. Rather
as Figs. \ref{etp-6} and \ref{etp-0} show, the matter plus vacuum
contributions to the renormalized energy and pressure satisfy the de
Sitter equation of state, $\varepsilon = -p$ at late times. In this
free field theory there is no scattering between the created
particles, and a `normal' equation of state cannot be established. The
form of the equation of state is expected to change as soon as
interactions are introduced, no matter how weak, since the particles
have a finite density and an unlimited interval of time to interact.

Next we consider the massless cases ($m=0$) for both conformal ($\xi =
{1/ 6}$) and minimal ($\xi =0$) couplings. As already remarked, in the
conformal case there is no particle creation at all, so the energy and
pressure remain strictly constant and agree with that of the
Bunch-Davies vacuum~\cite{bunch-davies} given by
\begin{eqnarray}
\langle T_{ab}\rangle^{\rm B-D}\left(m=0,\xi=\frac{1}{6}\right)&=&
- \frac{g_{ab}}{64 \pi^2}\frac{R^2}{2160} = 
- \frac{\bar H^4}{960\pi^2}g_{ab} \; ,
\end{eqnarray}
which corresponds to the value of the anomalous trace. 

\begin{figure}
\epsfxsize=15cm
\epsfysize=8cm
\centerline{\epsfbox{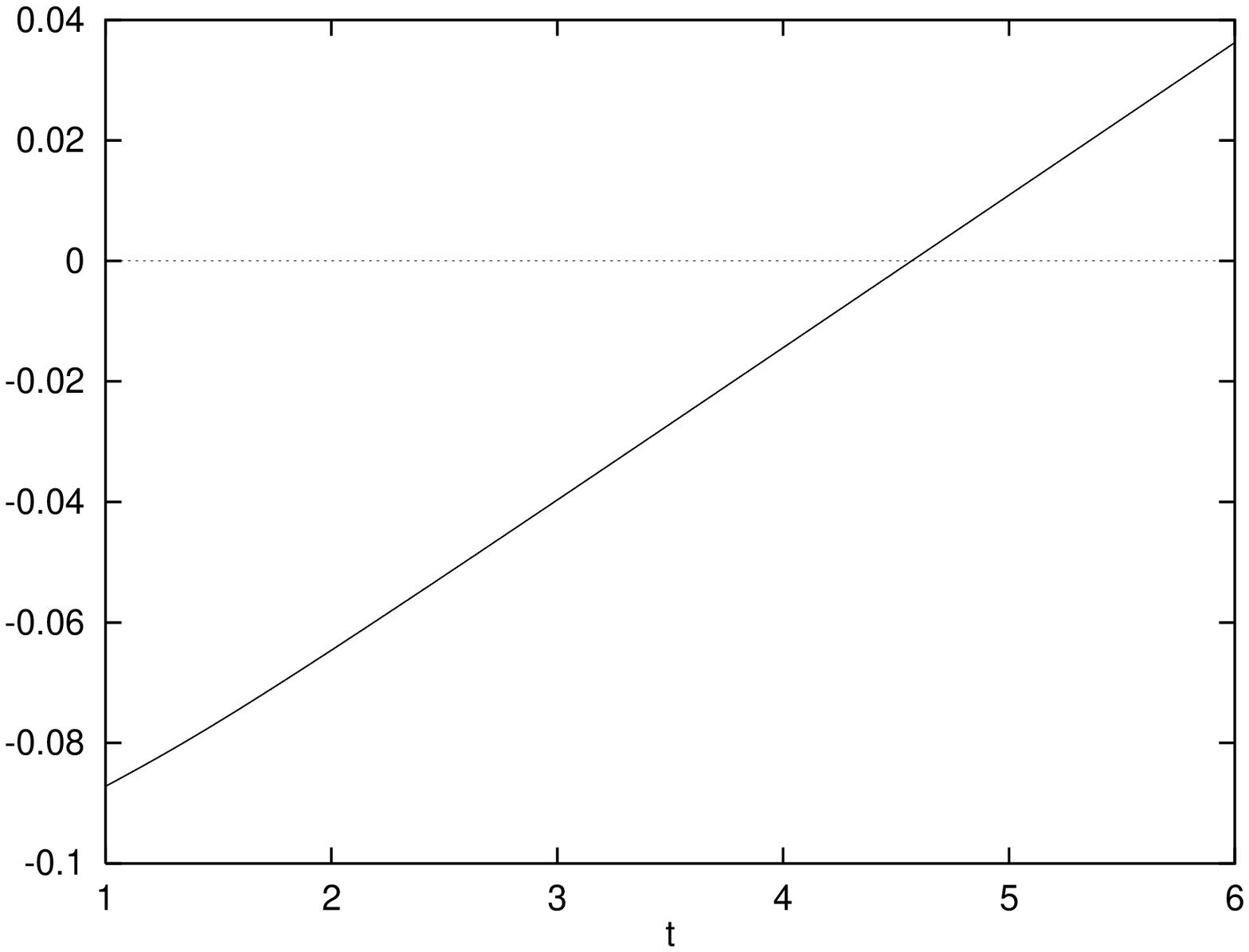}}
\vspace{0.25cm}
\caption{Renormalized value of $\langle \Phi^2 \rangle_R$  as a function
 of comoving time for $m=0$, $\xi=0$, and $t_0=1.0$.}
\label{phi2}
\vspace{0.25cm}
\epsfxsize=15cm
\epsfysize=8cm
\centerline{\epsfbox{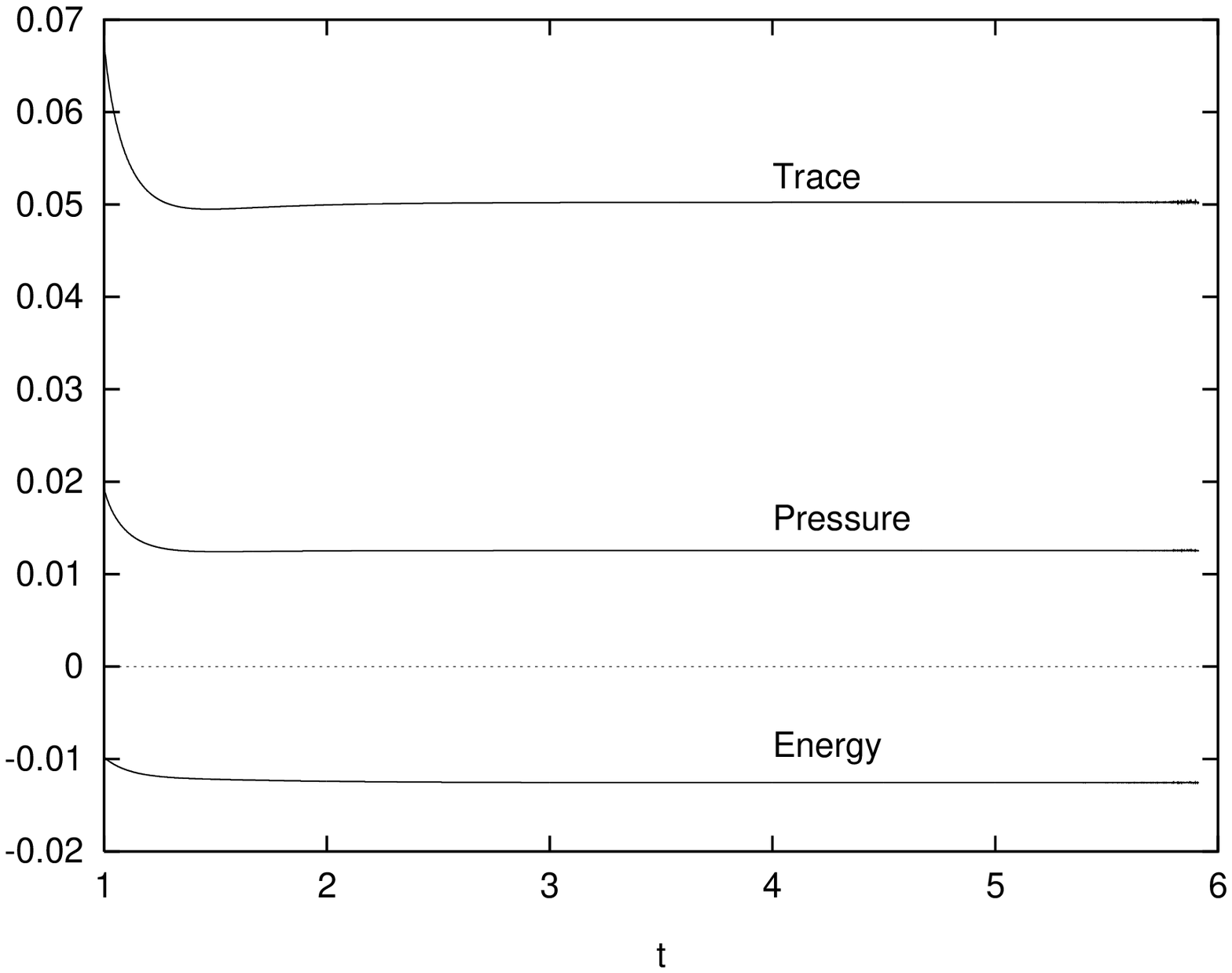}}
\vspace{0.25cm}
\caption{Energy, trace, and pressure as functions of comoving time
for $m=0$, $\xi=0$, and $t_0=1.0$.}
\label{tensor00}
\end{figure}

The results for the massless, minimally coupled field are shown in
Figs. \ref{phi2} and \ref{tensor00}. We point out the fact that the
renormalized value of $\langle \Phi^2 \rangle_R$ grows linearly in
time with a slope of $\frac{{\bar H}^3}{4 \pi^2}$, that is independent
of the initial conditions~\cite{smallm,habib}.
Notice that for the massless, minimally coupled case the late time
behavior of the renormalized energy, trace, and pressure are
determined by the value given in reference~\cite{folacci}, and it does
{\it not} go to the value calculated in reference~\cite{bunch-davies}.
This difference can be understood as a finite additional term that
must be added to the Bunch-Davies result when the spatially
homogeneous $k=1$ mode is handled properly.

It is interesting to study the case of the exactly massless, minimally
coupled scalar field for a wider variety of initial conditions.  If a
non-zero number of particles $N_k$ in a finite number of modes is
added, the asymptotic values of the renormalized $\langle
T_{ab}\rangle$ still approaches the de Sitter invariant
value~\cite{folacci}
\begin{eqnarray}
\langle  T_{ab}\rangle_{_R} \rightarrow {119 \bar H^4\over 960 \pi^2} 
g_{ab}\,, 
\end{eqnarray}
even though the initial state and transient behavior in those cases
(non-vacuum initial conditions) are not de Sitter invariant.

The fact that in all the above cases, even for de Sitter non-invariant
initial conditions and even for the massless, minimally coupled field,
the total renormalized energy-momentum tensor goes to the de Sitter
form, $\varepsilon \rightarrow -p \rightarrow $ constant, is an
important general conclusion of our study. Qualitatively, one can
understand this result in the following way. At late times, $t \gg
\bar H^{-1}$, $a(t)$ grows exponentially in de Sitter spacetime, and a
very large number of modes have wavenumber $k \ll a$, {\it i.e.,}
physical wavelengths that have been redshifted out of the de Sitter
horizon and frozen out. Those wavenumbers for which $k$ is still
greater than $a(t)$ behave like adiabatic vacuum modes and are removed
by the subtraction procedure.  Hence the sums (integrals) over $k$ in
the renormalized energy-momentum tensor are cut off at $k_{\rm max}$
of order $a(t)$. Since the initial conditions differ from the de
Sitter invariant vacuum data only for some finite set of wavenumbers
$k \le k_0$ (say), these small wavenumbers make a decreasingly small
contribution to the mode sum, which is controlled by the wide band of
wavenumbers for which $k_0 \ll k < a(t)$. The summand (integrand) in
this wide $k$ band appearing in the energy density, pressure or trace
is a function only of physical wavenumber $k/a$ (since it is
independent of the initial data). Examining the behavior of the mode
functions for late times and expanding the integrand in $\varepsilon$
or $p$ in powers of physical momentum it is easy to see that
\begin{eqnarray} {1\over a^3(t)}\sum_{k=k_0}^{a(t)} k^2 \left({a(t)\over
k}\right)^n  &\simeq &
a^{n-3}(t) \int_{k_0}^{a(t)} k^{2-n} {\mathrm{d}} k = 
1 - \left({k_0\over a(t)}\right)^{3-n}
\; 
\rightarrow 1 \; , \qquad {\rm for} \qquad n < 3
\; .
\label{asympn}
\end{eqnarray}
Hence all terms in the summand which do not grow faster than $n=3$ for
large $a(t)$ yield a constant contribution to $\langle T_{ab}\rangle$
in the late time limit, independent of the initial data.  The largest
power $n$ which can occur in the mode sums for $\varepsilon$ and $p$
is controlled by the behavior of the mode functions $f_k$ at late
times. Referring back to Eqs. (\ref{modefns}) and (\ref{bigom})
it is clear that the late time behavior of $f_k(t)$ is oscillatory,
$\exp (\pm i \gamma \bar H t)$ for $\gamma^2 \ge 0$ but exponential,
$\exp (\pm \vert\gamma\vert \bar H t)$ for $\gamma^2 < 0$. Since
$\langle T_{ab} \rangle$ is bilinear in the mode functions the power
$n$ can be as large as $2 \vert\gamma \vert$ for $\gamma$ pure
imaginary. Provided $\gamma^2 > - {9\over 4}$, all $n<3$, and the
exponential behavior cannot overcome the normal kinematic
redshift $a^{-3}$, there are no extreme infrared singular terms 
in the summand, and we conclude from (\ref{asympn}) that
both $\varepsilon$ and $p$ should tend asymptotically to a constant,
independent of the initial data at late times. By the covariant conservation 
equation (\ref{cons}) for the renormalized $\langle T_{ab}\rangle$ 
this implies the de Sitter equation of state $\varepsilon\rightarrow -p$ 
at late times, which is precisely what the numerical results show.

Conversely, if $\gamma^2 < - {9\over 4}$, we should expect the
integrals for the finite renormalized energy density and pressure of
the scalar field to contain terms with $n = 2 \vert \gamma\vert >3$
and to grow with the scale factor like $a^{2 \vert \gamma\vert -
3}$~\cite{sahni-habib}. Details of this case and the analytic
calculation of this behavior will be presented
elsewhere~\cite{smallm}.
 
The case $\gamma^2 = - {9\over 4}$ is marginal. For this value of
$\gamma^2$ the power $n=3$ is possible (but not required) in the mode
sum. If present, it leads to a logarithm, $\log a \rightarrow \bar H t$ and
linear growth in cosmic time for the components of the energy-momentum
tensor. One has to examine the energy-momentum tensor in more detail
to determine if such logarithmic behavior exists or not, as a general
argument based on the asymptotics of the mode equation (\ref{modefns})
will not suffice. In the adiabatic `vacuum' and matter contributions
separately we observe this linear growth for the massless, minimally
coupled scalar. This behavior of the `vacuum' component is easily
understood from the asymptotic forms of the functions $W_k$ and $V_k$
in (\ref{mmc}) in the massless, minimally coupled case, since they
imply that
\begin{equation}
\varepsilon^{\cal N}_k \rightarrow \left({a\over k}\right)^3 \, ,
\end{equation}
for $k \ll a$. This is the marginal $n=3$ power behavior which leads
to a linear growth (for the energy density of the vacuum contribution)
in comoving time with slope $(8 \pi^2)^{-1}$.  However, it turns out
that for the exactly massless, minimally coupled case ($m^2 = \xi =
0$) the complete, vacuum {\it plus} particle, renormalized energy and
pressure do not have such $n=3$ logarithmic divergences, as our
numerical studies show. In this case there is an exactly compensating
growth in the matter contribution which cancels the vacuum piece. This
cancellation is a consequence of the exact field translational
symmetry ($\Phi \to
\Phi + {\rm constant}$ )
in the massless, minimally coupled case which prevents a term
proportional to $\langle \Phi^2\rangle$ from appearing in $\langle
T_{ab}\rangle$.

\begin{figure}
\epsfxsize=15cm
\epsfysize=8cm
\centerline{\epsfbox{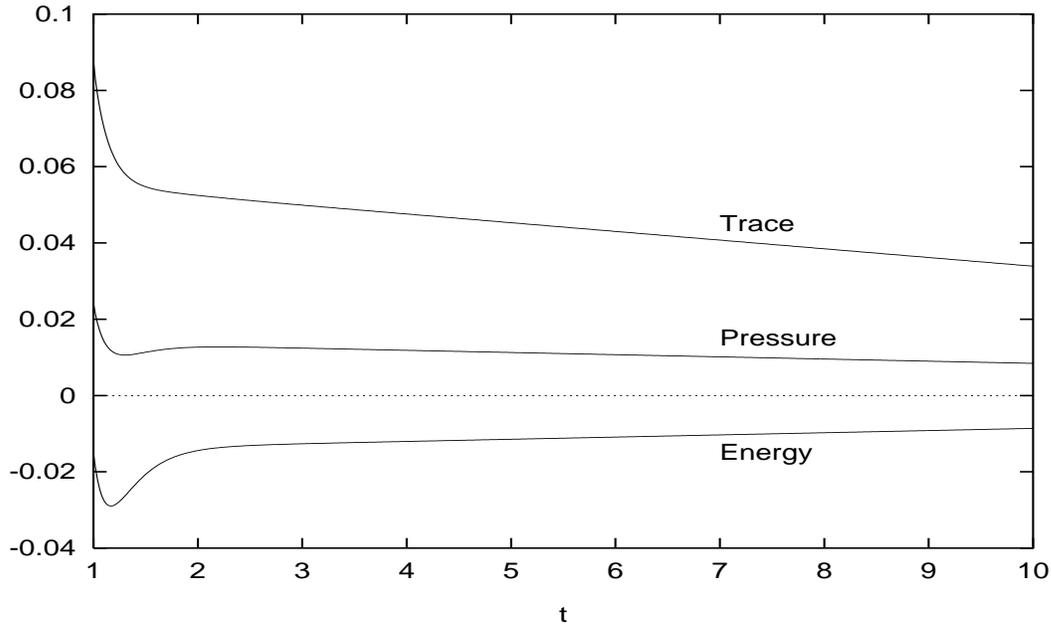}}
\vspace{0.25cm}
\caption{Energy, trace, and pressure as functions of comoving time
for $m=0.3$, $\xi=-0.0075$, and $t_0=1.0$.}
\label{nu32}
\end{figure}

Next we consider $\gamma^2 = - {9\over 4}$, {\it i.e.,} $m^2 + 12\xi
\bar H^2= 0$, with $m^2$ and $\xi$ separately different from zero,
which has the same mode equation but no such symmetry under field
translations. Because the energy-momentum tensor (\ref{ttij}) depends
on $m^2$ and $\xi$ in distinctly different ways, in these cases there
is a generic term proportional to $\langle \Phi^2\rangle$ in
$\varepsilon$ and $p$, and hence the exact cancellation of the linear
growth in $t$ between vacuum and particle contributions no longer
occurs. The {\it total} energy-momentum tensor now grows
logarithmically with $a(t)$ or linearly with $t$ in de Sitter
space. This behavior is illustrated in Fig. \ref{nu32}.
The coefficients of this linear growth can be determined analytically
from the known behavior of $\langle \Phi^2\rangle$, namely
\begin{eqnarray}
\varepsilon &\rightarrow & - {3 \xi \bar H^5 \over 4\pi^2} t 
+ \varepsilon_0
\,,\nonumber\\
p &\rightarrow & + {3 \xi \bar H^5 \over 4\pi^2} t 
+ p_0\,,
\nonumber\\
\langle T \rangle  &\rightarrow & + {12 \xi \bar H^5 \over 4\pi^2} t 
+ T_0\,,
\end{eqnarray}
with the constants $\varepsilon_0$ and $p_0$ constrained by the
conservation equation (\ref{cons}) to obey
\begin{equation}
\varepsilon_0 + p_0 = \xi {\bar H^4\over 4\pi^2}\,,
\end{equation}
and the constant $T_0$ determined by the equation
\begin{equation}
T_0= 3 p_0 -\varepsilon_0\,.
\end{equation}
Hence for $\xi > 0$ (and $m^2 < 0$) this linear growth decreases the
effective cosmological constant over time, whereas for $\xi < 0$ (and
$m^2 > 0$) it increases it. In either case the backreaction of the
scalar quantum field cannot be neglected at late times. It remains to
be seen whether this linearly growing behavior carries over to the
physically more relevant case of one-loop metric fluctuations
themselves, and if so with which sign.

Finally, it is interesting to study the case of minimal coupling and
non-vanishing but small real mass.  In this case there are two time
scales: a short time scale on which the Allen-Folacci vacuum is
approached, and a longer time scale controlled by $1/m$ on which there
is a smooth transition to the Bunch-Davies vacuum. Details of this
case and a reliable analytic calculation of the behavior observed will
be presented elsewhere~\cite{smallm}.

\section{Conclusions}

We have presented a general method to study the initial value problem
of a quantum field in a FRW spacetime. The two basic ingredients are:
an initial adiabatic state (of order four), and an adiabatic number
basis, with respect to which we define a particle number. This basis
allows one to write the bare energy-momentum tensor as a sum of two
contributions: a vacuum polarization term, and a particle production
term. The advantage of the basis presented here is that the
non-universal power law divergences of $\langle T_{ab}\rangle$ are
isolated in the vacuum polarization piece, and the matter contribution
displays physically reasonable behavior, as evidenced by the particle
creation process in de Sitter space.  We have implemented a numerical
scheme incorporating this theoretical approach which preserves the
covariant conservation of $\langle T_{ab}\rangle$, and yields a
renormalized $\langle T_{ab}\rangle$ which is cutoff independent.

The numerical results for a de Sitter background can be summarized as
follows. For the case of large mass ($\gamma \ge 0$) particle
production takes place at horizon crossing, and the late time behavior
of the renormalized energy-momentum tensor agrees with that of the
Bunch-Davies vacuum. In the conformally invariant case ($m=0,\xi=1/6$)
there is no particle production, and the renormalized value of the
energy-momentum tensor corresponds to the well-known trace
anomaly. The massless minimally coupled field ($m=0,\xi=0$) appears to
be a marginal case. Non-de Sitter invariant initial conditions are
driven to the Allen-Folacci vacuum, and not to the Bunch-Davies
vacuum. In none of the above cases does $\langle T_{ab}\rangle$ vanish
for late times, that is, it is not red-shifted away due to the
expansion of the universe. Backreaction effects, though not required
by this behavior cannot be ruled out when higher order corrections due
to self-interactions or fluctuations in the background metric are
included. Finally, for $m^2 + \xi R = 0$, but $m^2$ and $\xi$
individually different from zero, the total energy density, trace, and
pressure of a free scalar field grow without bound, logarithmically
with the scale factor. In this case it is certain that backreaction
effects must be included self-consistently, since the energy density
and pressure must eventually grow large enough to influence the
background geometry and invalidate the test field approximation.

We believe that these results merit a fuller investigation of the
effects of self-interactions and quantized gravitational fluctuations
in a complete backreaction calculation. With respect to the latter, an
interesting extension of the present study of free scalar field theory
would be to the second order energy-momentum tensor of quantized
metric fluctuations themselves. Barring a special cancellation which
occurs in the $\xi = m =0$ scalar case, we may expect that this
energy-momentum tensor would contain terms which grow logarithmically
with the scale factor, and dynamically influence the evolution of the
mean geometry away from de Sitter space, as has been suggested by
general arguments some time ago~\cite{mazagon}.  The generalization to
fully self-interacting fields is another interesting direction to
pursue, since one would again expect the interactions to alter the de
Sitter equation of state found in our study of the asymptotic behavior
of massive fields. The formalism presented in this paper is intended
to serve as a foundation from which the detailed analyses of these
issues can begin, and the general backreaction problem in cosmological
spacetimes with arbitrary finite adiabatic initial data can be posed
and solved numerically. These extensions, applications to specific
cosmological models, and the generation of density perturbations in
the early universe are under current study.

\section*{Acknowledgments}

The authors wish to express their gratitude to Paul Anderson for his
incisive comments on the manuscript and endless hours of fruitful
discussion.  Numerical computations were performed on the T3E at the
National Energy Research Scientific Computing Center (NERSC) at
Lawrence Berkeley National Laboratory (LBNL).

\appendix

\section{Adiabatic Expansion in Conformal and Comoving Time}

Because of the loss of explicit covariance under time
reparameterizations in the adiabatic regularization procedure, we
establish in this Appendix the relationship between the adiabatic
expansions in two different, commonly used time coordinates, namely
conformal time $\eta$ and the comoving time $t$ employed in the
present work. We will show that the two adiabatic expansions (comoving
and conformal) give identical results for physical quantities such as
$\langle T_{ab}\rangle$ to any order in the expansion.

The relation between the two time coordinates is given by
(\ref{conftim}) of the text. We denote a derivative with respect to
comoving time by a dot, and a derivative with respect to conformal
time by a prime. If we define the conformal mode functions by the
relation,
\begin{eqnarray}
g_k(\eta) = a^{-{1\over 2}}(t) f_k (t)
\; ,
\end{eqnarray}
then the mode equation (\ref{modefns}) in comoving time implies the
mode equation,
\begin{eqnarray}
g''_k(\eta) + \varpi^2_k (\eta) g_k (\eta)=0
\,,
\label{etamode}
\end{eqnarray}
with
\begin{eqnarray}
\varpi^2_k (\eta)= C(\eta)\omega_k^2
+\left( \xi - \frac{1}{6}\right) C(\eta) R(\eta) = a^2\Omega_k^2 + 
a^2 \left({\dot H\over 2} + {H^2\over 4}\right)
\,,
\label{Omom}
\end{eqnarray}
in conformal time. Hence the zeroth order adiabatic
frequency for the adiabatic expansion in conformal time is $a$ times
the zeroth adiabatic frequency for the adiabatic
expansion in comoving time, $\omega_k$. We will now show that
this feature holds to any adiabatic order.

Since the mode equations, (\ref{etamode}) or (\ref{modefns}), both
have the standard form of a time dependent harmonic oscillator
equation, the adiabatic ansatz,
\begin{eqnarray}
g_k(\eta) \equiv  {1 \over {{[2 \Upsilon_k(\eta)]}^{1/2}}} 
{\rm exp}\left[-i{\int^{\eta} {\mathrm{d}} \eta' \Upsilon_k(\eta')}\right] 
\,,
\end{eqnarray}
leads to an equation for $\Upsilon_k(\eta)$ in conformal time,
\begin{eqnarray}
\Upsilon^2_k = \varpi^2_k + {3 \over 4}
\frac{1}{\Upsilon_k^2} {\left(\frac{{\mathrm{d}} \Upsilon_k}
{{\mathrm{d}}\eta} \right)}^2
- {1 \over 2} \frac{1}{\Upsilon_k}
\frac{{\mathrm{d}}^2 \Upsilon_k}{{\mathrm{d}}\eta^2}
\,,
\label{adbeta}
\end{eqnarray}
which is exactly of the same form as that in comoving time, 
upon replacing $\Upsilon_k$ by $W_k$, and $\varpi_k$ by $\Omega_k$,
{\it i.e.,}
\begin{eqnarray}
W^2_k = \Omega^2_k + {3 \over 4}
\frac{1}{W_k^2} {\left(\frac{{\mathrm{d}} W_k}{{\mathrm{d}}t} \right)}^2
- {1 \over 2} \frac{1}{W_k}
\frac{{\mathrm{d}}^2 W_k}{{\mathrm{d}}t^2}\,.
\end{eqnarray}
To zeroth adiabatic order we have established already that
\begin{eqnarray}
\Upsilon_k^{(0)\,2} = C W_k^{(0)\,2} = C \omega_k^2\,.
\end{eqnarray}
Let us now assume that this result is true at adiabatic order $N-1$,
{\it i.e.,}
\begin{eqnarray}
\Upsilon_k^{(N-1)2}(\eta)= C(t)W_k^{(N-1)2}(t)\,.
\end{eqnarray}
and iterate to order $N$. Since
\begin{eqnarray}
\frac{{\mathrm{d}} \Upsilon_k^{(N-1)}}{{\mathrm{d}}\eta}&=&
a^2\left[
H W^{(N-1)}_k+\dot W_k^{(N-1)}\right]
\,,
\nonumber \\
\frac{{\mathrm{d}}^2 \Upsilon_k^{(N-1)}}{{\mathrm{d}}\eta^2}&=&
a^3\left[2H^2 W^{(N-1)}_k+\dot H W^{(N-1)}_k
+3 H \dot W_k^{(N-1)} +\ddot W_k^{(N-1)}\right]\,,
\end{eqnarray}
substitution into the right side of (\ref{adbeta}) and using
(\ref{Omom}) gives the desired iteration to order $N$, namely,
\begin{eqnarray}
\Upsilon^{(N)2}_k &=& \varpi^2_k + {3 \over 4}
\frac{1}{\Upsilon_k^{(N-1)2}} {\left(\frac{{\mathrm{d}}
\Upsilon_k^{(N-1)}}{{\mathrm{d}}\eta}
\right)}^2 - {1 \over 2} \frac{1}{\Upsilon_k^{(N-1)}}
\frac{{\mathrm{d}}^2 \Upsilon_k^{(N-1)}}{{\mathrm{d}}\eta^2}\nonumber\\
&=& a^2\left[\Omega_k^2 + {3 \over 4}
\frac{1}{W_k^{(N-1)2}} {\left(\frac{{\mathrm{d}} W_k^{(N-1)}}{{\mathrm{d}}t}
\right)}^2 - {1\over 2} \frac{1}{W_k^{(N-1)}}
\frac{{\mathrm{d}}^2 W_k^{(N-1)}}{{\mathrm{d}}t^2}\right]\nonumber\\
&=& a^2 W_k^{(N)2}\,.
\end{eqnarray}
Hence by induction we have proven that to any adiabatic order
\begin{eqnarray}
\Upsilon^{(N)}_k (\eta) = a(t) W_k^{(N)}(t)\,.
\end{eqnarray}
Moreover it is clear that this argument has relied only on the form of
the mode equation (\ref{modefns}) or (\ref{etamode}), and their
corresponding WKB definition of the frequencies $W_k$ and
$\Upsilon_k$. Since under an {\it arbitrary} differentiable
transformation of time coordinates,
\begin{eqnarray}
{\mathrm{d}}t = b(t') {\mathrm{d}}t'
\; ,
\label{ttrans}
\end{eqnarray}
the mode functions can be rescaled,
\begin{eqnarray}
f_k(t) = b^{-{1\over 2}}(t') g_k(t')
\; ,
\end{eqnarray}
in order for the first $t'$ derivative terms in $\ddot f_k$ 
to be eliminated, the mode equation in the new time variable
can be put in the standard form,
\begin{eqnarray}
g''_k(t') + \varpi^2_k (t') g_k (t')&=&0
\; ,
\,\qquad {\rm with}\qquad
\varpi^2_k (t') = b^2 \Omega_k^2 -{3\over 4} \left({b'\over b}\right)^2+
{b''\over 2 b}\,,  
\end{eqnarray}
where now the prime denotes differentiation with respect to $t'$.
Once in this form the adiabatic ansatz (\ref{adbeta}) may be
inserted and the previous inductive argument may be repeated
to yield
\begin{eqnarray}
\Upsilon^{(N)}_k (t') = b W_k^{(N)}(t)\,.
\end{eqnarray}
for the arbitrary reparameterization of the time coordinate,
$t \rightarrow t'(t)$.

An immediate corollary of this result is that any quantity
such as the energy density and pressure, developed up to
a given adiabatic order $N$ in the comoving time coordinate $t$
corresponds exactly to the same quantity developed up
to the same adiabatic order $N$ in any other time coordinate $t'$.
Since $T_{ab}$ transforms as a tensor, the energy density and
pressure transform as
\begin{eqnarray}
\varepsilon &\rightarrow &\varepsilon' = b^2 \varepsilon\,,\nonumber\\
p &\rightarrow & p' = p
\; ,
\end{eqnarray}
under (\ref{ttrans}). Hence up to the overall factor of $b^2$ in the
first of these relations, the energy density and pressure in either
time coordinate are numerically equal at any order of the adiabatic
expansion, and the renormalization procedure of subtracting the
expansion up to adiabatic order two or four from the bare quantities
is the {\it same} procedure in any time coordinate. The case $t' =
\eta$ and $b=a$, relating conformal time to comoving time, is just a
special case of this much more general result, so that the conformal
adiabatic expansion which has been given, for example,
in~\cite{anderson-parker,Bunch} is exactly equivalent to the
corresponding expansion in comoving time given by Eqns.
(\ref{adbtwo}) and (\ref{adbfour}) of the text, as may be checked by
explicit comparison of the two sets of expressions.  Hence the
subtraction of the adiabatic order two expressions (\ref{eTtwo}) from
the bare energy density and pressure corresponds exactly to the same
subtraction in conformal (or any other) time coordinate which leaves
the spatial sections unchanged. Finally, since the fourth order
logarithmic cutoff dependence of $\langle T_{ab}\rangle$ is
proportional to the geometric tensor $^{(1)}H_{ab}$, this remaining
cutoff dependence and the argument (given in the text) for how to
handle it in backreaction calculations, applies equally well in any
time coordinate.

\end{document}